\documentclass[useAMS,usenatbib]{mn2e}
\usepackage{graphicx}
\usepackage{epstopdf}
\usepackage{amsmath}
\usepackage{amssymb}
\usepackage{aas_macros}
\usepackage{multirow}
\usepackage{hyperref}
\usepackage[svgnames]{xcolor}

\def \Mpc {\, \mathrm{Mpc} \,}
\def \Mpch {\, \mathrm{Mpc}\,h^{-1} \,}
\def \Hunit {\, \mathrm{km} \, \mathrm{s}^{-1}\mathrm{Mpc}^{-1} \,}

\begin{document}

\title[The distance-redshift relation from the BAO of galaxy
  clusters]{Measuring the distance-redshift relation with the baryon
  acoustic oscillations of galaxy clusters}

\author[Veropalumbo, et al.]
       {A. Veropalumbo$^1$\thanks{E-mail:
           alfonso.veropalumbo@unibo.it}, F. Marulli$^{1,2,3}$,
         L. Moscardini$^{1,2,3}$, M. Moresco$^{1,2}$ and A. Cimatti$^1$
         \\ $^1$Dipartimento di Fisica e Astronomia, Universit\`a di
         Bologna, Alma Mater Studiorum, viale Berti Pichat 6/2, I-40127 Bologna,
         Italy\\ $^2$INAF - Osservatorio Astronomico di Bologna, via
         Ranzani 1, I-40127 Bologna, Italy\\ $^3$INFN - Sezione di
         Bologna, viale Berti Pichat 6/2, I-40127 Bologna, Italy}
       
\pagerange{\pageref{firstpage}--\pageref{lastpage}}
       
\maketitle
\label{firstpage}

\begin{abstract}
  We analyse the largest spectroscopic samples of galaxy clusters to
  date, and provide observational constraints on the distance-redshift
  relation from baryon acoustic oscillations. The cluster samples
  considered in this work have been extracted from the Sloan Digital Sky
  Survey at three median redshifts, $z=0.2$, $z=0.3$, and $z=0.5$. The
  number of objects is $12910$, $42215$, and $11816$, respectively. We
  detect the peak of baryon acoustic oscillations  for all the three samples. 
  The derived distance constraints are: 
    $r_s/D_V(z=0.2)=0.18 \pm 0.01$, $r_s/D_V(z=0.3)=0.124 \pm 0.004$
    and $r_s/D_V(z=0.5)=0.080 \pm 0.002$. 
  Combining these measurements, we obtain robust constraints on cosmological
  parameters. Our results are in agreement with the standard $\Lambda$
  cold dark matter model. Specifically, we constrain the Hubble
  constant in a $\Lambda$CDM model, $H_0 = 64_{-9}^{+14} \, \Hunit$, 
  the density of curvature energy, in the $o\Lambda$CDM context, 
  $\Omega_K = -0.015_{-0.36}^{+0.34}$, and finally the parameter of the dark energy 
  equation of state in the $ow$CDM case, $w = -1.01_{-0.44}^{+0.44}$. 
  This is the first time the distance-redshift relation has been constrained 
  using only the peak of baryon acoustic oscillations of galaxy clusters.
\end{abstract}



\section{Introduction}
\label{sec:intro}
Galaxy clusters play a leading role in both present and planned
cosmological investigations \citep[see e.g][]{paper:allen2011}.  They
represent the biggest collapsed structure of the Universe, sitting on
top of the highest peaks of the dark matter density field. The
possibility of modelling their statistical properties as a function of
cosmological parameters, combined with the capability of measuring
their basic properties, such as the mass, with relative simplicity
with respect to other astrophysical objects, makes them optimal
tracers of the large scale structure of the Universe. Recent works
based on multiple wavelength observations have already reached
important goals in defining the knowledge of the Universe today. A
powerful strategy is to extrapolate cosmological information from
second-order statistics, i.e. two-point correlation function (2PCF) or
power spectrum, of galaxy clusters as a standalone
\citep{paper:veropalumbo2014} or in a joint analysis with mass
function measurements \citep[see][and reference
  therein]{paper:mana2013, paper:sartoris2015}.  Clustering contains
plenty of cosmological information at different scales. Among all, the
baryon acoustic oscillation (BAO) peak in the 2PCF is currently one of
the most important cosmological probe \citep[see
  e.g.][]{paper:eisenstein1998}.  BAO is an oscillation pattern in the
matter power spectrum, counterpart of the acoustic oscillations seen
in the angular power spectrum of the cosmic microwave background.  It
has been generated by sound waves in the baryon-photon fluid before
recombination, as a consequence of the presence of anisotropies in the
dark matter distribution.  The typical length scale associated with
this feature is the sound horizon, $r_s \approx 150 \Mpc$, that is the
maximum distance a sound wave can propagate before decoupling ($z_D
\simeq 1100$) given its sound speed.  This scale remains imprinted
after decoupling in the distribution of matter and can be measured
from clustering of different tracers of the underlying dark matter
distribution. This capability makes the BAO a convenient probe to
exploit the technique of the {\em standard ruler}, and to map the
distance-redshift relation to get cosmological constraints.  An
increasing number of distance measures at local redshift ($z<1$) has
been obtained in the last years, thanks to wide surveys of galaxies,
such as 6dFGS \citep{paper:beutler2011}, BOSS
\citep{paper:anderson2014, paper:cuesta2015}, and WiggleZ
\citep{paper:blake2011, paper:kazin2014}. BAO as a standard ruler can
also provide important information when extracting full-shape
constraints from the 2PCF and power spectrum \citep[see
  e.g.][]{paper:sanchez2013}. Moreover, BAO can also be detected in
the clustering pattern of other tracers, such as Ly$\alpha$ emitters
\citep{paper:delubac2015} and galaxy clusters
\citep{paper:estrada2009, paper:hutsi2010, paper:hong2012,
  paper:veropalumbo2014}.

In this work we aim at obtaining a multi-redshift distance constraint by
measuring the BAO peak in the 2PCF of three spectroscopic samples of
galaxy clusters. In a forthcoming paper, we will use these cluster
catalogues to perform a joint analysis of the mass function and
clustering, to further tighten the cosmological constraints. This will
be an unique opportunity also to test these methodologies for future
analyses on next generation surveys, such as Euclid \citep{paper:laureijs2011,
  paper:amendola2013}, that will push our knowledge of the Universe to
an unprecedented level of precision.

The paper is organised as follows. We present our cluster samples in
\S~\ref{sec:data}, and show details on 2PCF measurements and cosmological
analyses in \S~\ref{sec:meas}. In \S~\ref{sec:res} we present our
results on clustering measurements, on the comparison with previous galaxy
studies and on the derived cosmological constraints.  Finally, in
\S~\ref{sec:conclusion} we draw our conclusions.


\section{Data}
\label{sec:data}
This section describes the data used for the clustering analyses. More
details on the methods exploited to detect the galaxy clusters and to
construct the spectroscopic samples can be found in
\citet{paper:whl2012} (WHL) and \citet{paper:veropalumbo2014}.


\subsection{Galaxy clusters}
The WHL catalogue consists of $132683$ galaxy clusters on a sky area
of $15000$ square degrees, in a redshift range of $0.05 < z < 0.8$.
The cluster identification is based of the friends-of
  friends procedure. This approach has been already exploited to find
  groups and clusters, using volume limited spectroscopic samples of
  galaxies \citep[see e.g.][]{paper:berlind2006,paper:tempel2014}, at
  low redshift ($z<0.2$). The WHL cluster sample extends the technique on
  photometric redshift samples of galaxies, allowing structure
  detections at higher redshift ($z<0.7$).  Structures are identified
starting from the positions of the observed galaxies in the
photometric sample of the Sloan Digital Sky Survey (SDSS) Data Release
8 (DR8) \citep{paper:sdssdr8}.  Clusters that enter the final
catalogue are those overdensities with richness $N_{200} \geq 8$, with
$N_{200}$ the number of members inside the estimated structure radius
$r_{200}$, and optical richness $R_{L_*} = L_{200}/L_*\geq 12$, being
$L_{200}$ the luminosity inside $r_{200}$ and $L_*$ the evolved
characteristic galaxy luminosity \citep{paper:blanton2003}.  The
position of the cluster center coincides with the angular position of
the brightest cluster member (BCG), while the photometric redshift is
assigned by averaging over all the member redshifts.  From weak
lensing scaling relations \citep[see e.g.][]{paper:wen2009,
  paper:covone2014}, the minimum mass for such structures results
$M_{200} > 0.6\times 10^{14} M_{\odot}$.


\subsection{Sample selection}
A precise estimate of the redshift is crucial when reconstructing
statistical properties of the large scale distribution of matter.
Large redshift errors, as in photometric redshift surveys, lead to
severe distortion effects that reflect in the 2PCF measurement,
complicating its analysis and cosmological interpretation \citep[see
  e.g.][]{paper:marulli2012, paper:sereno2014}.  In order to construct
spectroscopic cluster samples, we take advantage of the spectroscopic
data from the SDSS.  We focus on the SDSS DR7 \citep{paper:sdssdr7}
and on the final spectroscopic data release \citep{paper:sdssdr12}
from the Baryon Oscillation Spectroscopic Survey (BOSS), part of the
SDSS III program. We assign a redshift to a cluster if it has been
observed for its BCG.  Hereafter, we make no distinction between
galaxy clusters and BCGs because, by construction, the position of
each cluster is entirely determined by the coordinates of its BCG.  We
mask those clusters falling in bad photometry regions or near bright
stars. The masking procedure reduces the cluster
  sample by $5\%$.  We also exclude poor or failed spectroscopic
observations.  The total number of BCGs with a measured spectroscopic
redshift is $\approx 80000$.

Through this procedure we manage to obtain the {\em largest}
spectroscopic galaxy cluster catalogue to date.  Thanks to the high
abundance of tracers, we can split the catalogue in three subsamples
at different redshifts, according to the type of target each BCG is
assigned in the SDSS program. For details on the targeting selection,
we refer to \cite{paper:boss}.  We consider three types of targets:
the \textit{Main Galaxy Sample}, consisting of luminous
galaxies ($r<17.77$) at $ z< 0.3$; the \textit{LOWZ sample}, that
targets Luminous Red Galaxies up to a redshift $z<0.43$; the
\textit{CMASS sample}, focused on high-redshift galaxies in the range
$0.43 < z < 0.7$. 

The derived spectroscopic cluster catalogues are the following:
\begin{itemize}
  \item the Main Galaxy Cluster Sample (Main-GCS), consisting of
    $12910$ BCGs, part of the Main Galaxy sample in the north galactic
    cap;
  \item the LOWZ Galaxy Cluster Sample (LOWZ-GCG), with $42215$ BCGs
    in the LOWZ sample;
  \item the CMASS Galaxy Cluster Sample (CMASS-GCS), with $11816$ BCGs
    in the north galactic cap of the CMASS sample.
\end{itemize}

\begin{figure*}
\begin{center}
\includegraphics[width=\textwidth]{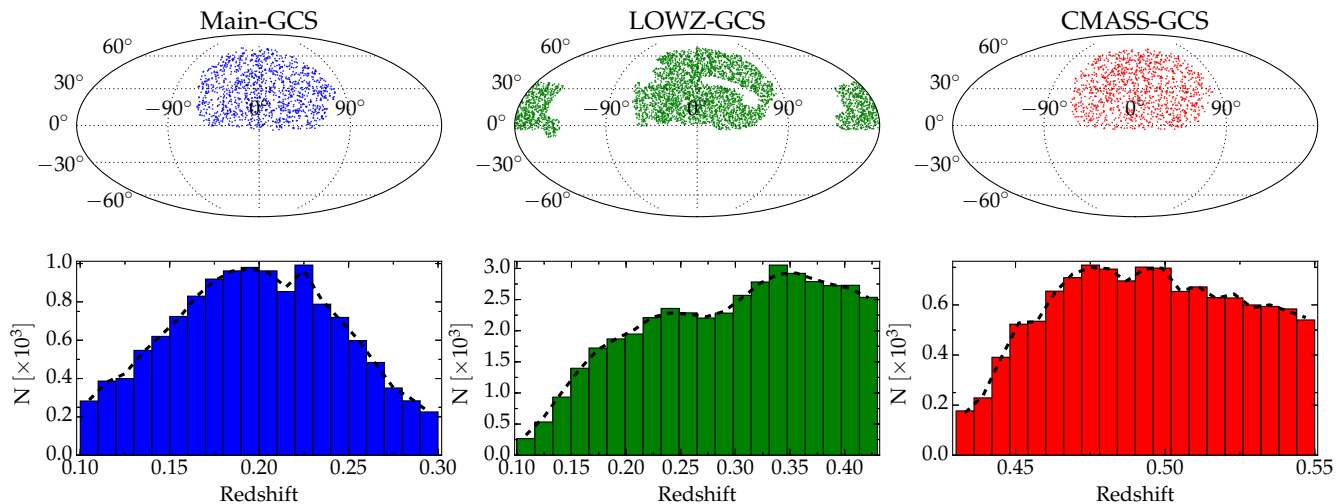}
\end{center}
\caption{The angular (top panels) and redshift distributions (bottom panels)
  of the three selected galaxy cluster catalogues: Main-GCS (blue),
  LOWZ-GCS (green) and CMASS-GCS (red). The black dashed curves in the
  bottom panels are the reconstructed redshift distributions used for
  the construction of random catalogues.}
\label{fig:zdist}
\end{figure*}

We restrict our redshift ranges to \textit{i}) $0.1\leq z \leq 0.3$
for the Main-GCS, \textit{ii}) $0.1\leq z \leq 0.43$ for the LOWZ-GCS,
and \textit{iii}) $0.43\leq z \leq 0.55$ for the CMASS-GCS.  
We have chosen the redshift cut at $z<0.43$ so that the CMASS-GCS and 
the LOWZ-GCS are independent samples. On the other hand, a significant 
fraction of clusters is in common between the Main-CGS and the LOWZ-CGS 
samples. We choose this sample splitting to maximize the number of clusters 
in each redshift bin and to simplify the creation of random catalogues. 
We discuss the covariance between samples in \S~\ref{sec:res}.  
Table \ref{tab:samples} reports the main properties of the selected samples, 
while in Fig.~\ref{fig:zdist} we show their angular (upper panels) 
and redshift distributions (lower panels). The LOWZ-CGS results to be 
the largest cluster sample ever used in a clustering analysis, almost two 
times larger than the one  used in \citet{paper:veropalumbo2014}. 

\begin{table*}
  \centering
    \caption{The main properties of the cluster samples used
    for the clustering analysis.  The bias has been obtained by
    modelling the projected correlation function in the scale range
    $5<r[\Mpch]<20$ (see \S \ref{subsec:bias}).}
  \begin{tabular}{l|c|c|c|c|c|c} 
    \hline
    \hline
    Sample Name & Number of clusters & Redshift Range & Median Redshift & bias \\
    \hline
    Main-GCS  & $12910$ & $0.1  \leq z \leq 0.3  $ & $0.20$ & $2.00 \pm 0.05 $ \\
    LOWZ-GCS  & $42115$ & $0.1  \leq z \leq 0.43 $ & $0.30$ & $2.42 \pm 0.02 $ \\
    CMASS-GCS & $11816$ & $0.43 \leq z \leq 0.55  $ & $0.50$ & $3.05 \pm 0.07$ \\
    \hline
    \hline
  \end{tabular}
  \label{tab:samples}
\end{table*}


\subsection{Weights}
\label{subsec:weight}
We apply a weight to each cluster to correct for mass and redshift
incompleteness (see WHL for further details).  The cluster samples
result to be $75\%$ complete for the minimum mass threshold, and up to
$100\%$ complete for $M_{200} > 2\times 10^{14} M_{\odot}$, at $z <
0.42$. The detection rate drops below $z > 0.42$; this explains the
difference in density between LOWZ-GCS and CMASS-GCS. Moreover, for
the CMASS-GCS sample, we take into account the dependence on seeing
and stellar density of targets on the celestial sphere, as introduced
in \citet{paper:anderson2012}, to obtain a more consistent estimate of
the 2PCF at large scales.  This weighting scheme
  lowers the 2PCF normalization by $< 10\%$. Thus, it can be
  considered as a minor effect considering the 2PCF uncertainites at
  the BAO scales.


\section{Clustering measurements}
\label{sec:meas}
In the following we describe all the steps concerning the estimate of
the 2PCF from the samples described above, and the distance
constraints obtained from the BAO fitting\footnote{To perform all the
  analyses presented in this paper we use the {\small CosmoBolognaLib}
  \citep{paper:marulli2015}, a large set of Open Source C++ libraries freely
  available at this link:
  http://apps.difa.unibo.it/files/people/federico.marulli3/.}


\subsection{The two-point correlation function}
\label{subsec:xi}
We measure the 2PCF using the \citet{paper:landy1993} estimator:
\begin{equation}
  \xi(s) = \frac{DD(s)+RR(s)-2DR(s)}{RR(s)} \, ,
  \label{eq:xiLS}
\end{equation}
where $DD(s)$, $RR(s)$ and $DR(s)$ are the data-data, random-random
and data-random normalized pairs counts, respectively, for a
separation bin $s\pm\mathrm{d}s/2$. We measure the 2PCF up to $200
\Mpch$ in bins of $8 \Mpch$, for the Main-GCS and the LOWZ-GCS, and in
bins of $10 \Mpch$ for the sparser CMASS-GCS sample.


\subsubsection{Geometrical distortions}
\label{subsec:geodist}
To estimate the comoving separations between object pairs, a
\emph{fiducial cosmology} has to be assumed. Indeed, the 3D 2PCF is
not a cosmology-independent quantity, and when constraining the BAO
peak we have to take into account the \emph{geometrical distortions}
introduced by a possible wrong assumption of the background cosmology.

As a fiducial cosmology, we assume a flat $\Lambda$ cold dark matter
(CDM) model with Hubble constant $H_0=68 \, \Hunit$, total matter
density parameter $\Omega_M=0.3$, baryon density parameter
$\Omega_b=0.045$, primordial spectral index $n_s = 1$, and 
matter power spectrum normalization corresponding to $\sigma_8 = 0.83$.


\subsubsection{Redshift-space distortions}
The measured redshifts do not contain only distance information, being
perturbed by the line-of-sight peculiar motions of the mass
tracers. This introduces the so-called \emph{dynamical distortions} in
the clustering pattern.  Specifically, both linear distortions, caused
by ordered large-scale flows, and non-linear distortions, caused by
random peculiar motions, are generally present.  While linear
distortions are always present, non-linear dynamical distortions have
a very minor impact on BCGs, compared to satellite galaxies,
since BCGs trace the bottom of the cluster potential
  wells.  We verified this important aspect by looking at the
  two-dimensional 2PCF of galaxies and galaxy clusters. The former
  shows a very clear signal of the so-called Fingers of God, due to
  the random motions of satellites in haloes. On the other hand, this
  signal is almost absent in the 2PCF of galaxy clusters
  \citep{paper:marulli2015b}.  Photometric redshift errors can be
fairly considered as a limiting case of peculiar motions.  We then
expect that non-linear distortions have limited effects on BCGs with
spectroscopic redshift \citep[see e.g.][and references
  therein]{paper:marulli2015b}.  This reflects in a sharper BAO signal
for this kind of tracers.  We discussed how much these effects impact
the galaxy cluster clustering in \citet{paper:veropalumbo2014}.


\subsubsection{Bias determination}
\label{subsec:bias}
To estimate the linear bias $b$ of our cluster samples, we use the
projected correlation function defined as follows:
\begin{equation}
  w_p(r_p) = \int_0^{\pi_{max}} \mathrm{d}\pi'\xi(r_p,\pi') \, ,
  \label{eq:xiproj}
\end{equation}
where $\xi(r_p,\pi)$ is the 2PCF measured in bins of perpendicular,
$r_p$, and parallel, $\pi$, separations with respect to the
line-of-sight.  Integrating along the direction parallel to the
line-of-sight allows us to approximately correct for redshift-space
distortions. In Eq.~\ref{eq:xiproj} we set $\pi_{max}=50 \Mpch$.

To obtain the bias, we model the projected correlation functions as
follows:
\begin{equation}
w_p(r_p) = b^2 \int_{r_p^2}^{\sqrt{\pi_{max}^2+r_p^2}} \mathrm{d}r
\frac{2 r \xi(r)}{\sqrt{r^2-r_p^2}} ,
\end{equation}
where $\xi(r)$ is the predicted matter power spectrum.
This measure allows us to construct a cluster 2PCF
  model, to be used in the covariance matrix estimate via lognormal
  mock catalogues (see \S~\ref{subsec:lnmock}).


\subsection{Random catalogues}
\label{subsec:random}
According to Eq. \ref{eq:xiLS}, a random sample has be provided to
correctly evaluate the 2PCF. A random catalogue contains the
information on the selection function of the data sample, that are
used to balance spurious effects affecting the pair counting.  The
selection function can be safely reproduced separating it into angular
and radial parts.  We generate random catalogues almost 20 times
larger than the reference cluster samples to limit shot noise effects.


\subsubsection{Angular mask}
We generate random points using publicly available survey footprints
\footnote{the SDSS DR7 window is available at
  \url{http://sdss.physics.nyu.edu/vagc/}; the BOSS survey footprint
  is available at \url{http://data.sdss3.org/sas/dr9/boss/lss/}.} and
the {\small MANGLE} software \citep{paper:swanson2008}. For the three
cluster samples we used the following masks:
\begin{itemize}
  \item Main-GCS: the SDSS DR7 survey footprint, using the window
    provided by the NYU-VAGC \citep{paper:blanton2005};
  \item Lowz-GCS: the BOSS survey footprint, excluding regions with
    bugged target selections (\small{IPOLY} $>10324$);
  \item CMASS-GCS: the Northern Galactic Cap of BOSS survey footprint.
\end{itemize}
We mask random points falling in the veto regions, as done for the data. 
 

\subsubsection{Cluster redshift distribution}
We consider two methods to assign redshifts to the random collections
of objects:
\begin{itemize}
  \item random extraction from the smoothed redshift distribution; the
    parameters involved are the redshift bin and the size of the
    Gaussian kernel;
  \item random assigning from the galaxy cluster redshifts \citep[see][] {paper:ross2012}.
\end{itemize}
We verified that all our results are robust independently of the
method and adopted parameters. In the following, we will show results
obtained with the first method, grouping the cluster redshift
distribution in $100$ bins and smoothing the redshift distribution
with a Gaussian kernel three times larger than the bin size.


\subsection{Covariance matrix}
\label{subsec:covmat}
The covariance matrix is a crucial ingredient for clustering analyses.
It measures the correlation between correlation function bins, and it is
defined as follows:
\begin{equation}
  C_{i,j} = \frac{1}{N-1} \sum_{k=1}^{N}(\xi^k_i-\hat{\xi_i}) (\xi^k_j-\hat{\xi_j}) \, ,
  \label{eq:jkcov}
\end{equation}
where the subscripts $i$ and $j$ run over spatial bins of the correlation function
and $k$ refers to the 2PCF of the $k^{th}$ of $N$ realizations;
$\hat{\xi}$ is the mean 2PCF of the $N$ realizations.

The covariance matrix can be directly estimated using mock catalogues
extracted from numerical simulations \citep[see
  e.g.][]{paper:anderson2014, paper:kazin2014}. However, this method
is very computational expensive if a large set of mocks have to be
created. Alternatively, different statistical techniques can be
exploited, that still provide fairly robust estimates of the
covariance matrix.  For our analysis, we consider the following
approaches to estimate the covariance matrix:
\begin{itemize}
\item two internal error estimators: jackknife and bootstrap
  \citep[see e.g.][]{paper:norberg2009};
\item one external error estimator, that exploits lognormal mocks
  generated from the density field.
\end{itemize}


\subsubsection{Internal errors}
The covariance matrix can be estimated by subsampling the original
catalogue and calculating the correlation function in all but one sub-
samples (jackknife), or in a random selection of them (bootstrap),
recursively.  We choose to subsample the observations in $50$ right
ascension-declination regions, that results in $50$ jackknife mock
realizations, while we extract $200$ times these subregions to exploit
the bootstrap resampling.


\subsubsection{External errors}
\label{subsec:lnmock}
We compare the internal error estimates described above with the ones
assessed through the lognormal density field technique
\citep{paper:coles1991}.  This method to infer the covariance matrix
has been already used by several authors for clustering analyses
\citep[see e.g.][]{paper:beutler2011, paper:blake2011,
  paper:chuang2014}.

We create the density field realizations using the redshift-space
monopole model:
\begin{equation}
  P_{model}(k) = b^2 \left( 1+\frac{2}{3}\beta+\frac{1}{5}\beta^2
  \right) P_{DM}(k) \, ,
  \label{eq:pkln}
\end{equation}
where $P_{DM}$ is the linear matter power spectrum obtained with the
{\small CAMB} software \citep{paper:lewis2002}, $b$ is the bias
constrained from the galaxy cluster projected correlation function at
the small scales, and $\beta$ is the ratio between the linear growth
rate function $f=\Omega_M(z)^{0.545}$, as predicted by General
Relativity, and the bias $b$. To measure $b$ we model the projected
correlation function as described in \S\ref{subsec:bias}. The
covariance matrix used for the fit is the one derived with the
sub-sampling approach. The estimated values of the bias 
and its standard deviation for each sample are reported in Table \ref{tab:samples}.

Density fields are generated in boxes large enough to contain the
survey volumes, with a regular grid of steps half the size of the bins
used to estimate the 2PCF. The survey selection function is taken into
account in the random catalogues. Once the mock clusters are extracted
according to the density distribution, the covariance matrix can be
directly estimated by measuring the 2PCF for each mock sample.


\subsection{Model}
In this section we describe the models used to derive distance
constraints from the BAO peak in the 2PCF.


\subsubsection{Fitting the 2PCF}
To extract cosmological constraints from the BAO peak, we adopt the
following widely used and robust model \citep[see][and reference
  therein] {paper:anderson2012}:
\begin{equation}
  \xi(r) \, = \, B^2\xi_{DM}(\alpha r) +A_0 + \frac{A_1}{r} + \frac{A_2}{r^2} \, ,
  \label{eq:ximodel}
\end{equation}
where $\xi_{DM}(r)$ is the dark matter 2PCF calculated assuming the fiducial
cosmology, $B$ factorises the difference between the dark matter 2PCF
and the cluster 2PCF, $\alpha$ is the parameter that contains the
distance information, and $A_0$, $A_1$ and $A_2$ are the parameters of
an additive polynomial used to marginalize over signals caused by
systematics not fully taken into account.

We adopt the \emph{de-wiggled} template for the dark matter power
spectrum, $P_{DM}$ \citep{paper:eisenstein2007}:
\begin{equation}
  P_{DM}(k) \, = \, \left[ P_{lin}(k) - P_{nw}(k) \right]  \mathrm{e}^{-k^2\Sigma_{NL}^2/2} + P_{nw}(k) \, ,
  \label{eq:pkmodel}
\end{equation}
where $P_{lin}$ is the linear power spectrum as provided by {\small
  CAMB} \citep{paper:lewis2002}, while $P_{nw}$ is the power spectrum
without the BAO feature, as obtained by the parametric formula of
\citet{paper:eisenstein1998}.  The parameter $\Sigma_{NL}$ controls
the smearing of the BAO, and it is left free to vary.  The model for
$\xi_{DM}(r)$ is then obtained by Fourier transforming the power
spectrum:
\begin{equation}
  \xi_{DM}(r) = \frac{1}{2\pi^2}\int \mathrm{d}k \, k^2 P_{DM}(k) \frac{\sin(kr)}{kr} \, .
  \label{eq:xipk}
\end{equation} 
To get marginalized constraints we populate the parameter space with
the Monte Carlo Markov Chain technique.  We adopt a Gaussian
likelihood, $ \propto \exp(-\chi^2/2)$, with $\chi^2$ defined as
follows:
\begin{equation}
  \chi^2 = \sum_{i=0}^n \sum_{j=0}^n (\xi_i-\xi^m(r_i))C^{-1}_{ij}(\xi_j-\xi^m_j) \, ,
  \label{eq:chi2}
\end{equation}
where $\xi_i$ is the 2PCF measured in the $i^{th}$ bin, $\xi^m_i$ is
the model in the same bin and $C^{-1}_{ij}$ is the inverted covariance matrix.

\begin{figure*}
  \begin{center}
  \includegraphics[width=\textwidth]{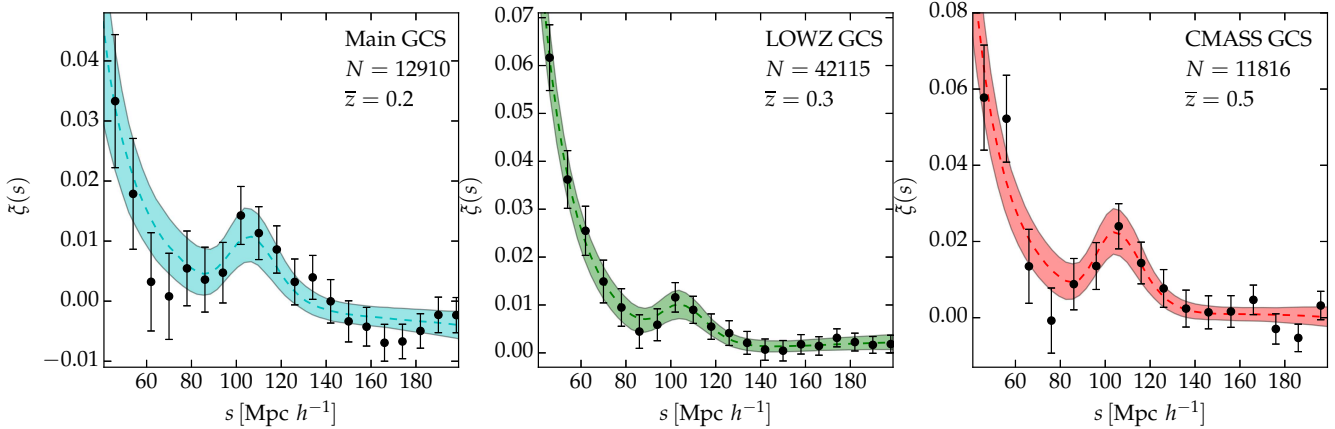}
  \end{center}
  \caption{The redshift-space 2PCF of galaxy clusters, respectively
   Main-GCS (left panel), LOWZ-CGS (central panel) and CMASS-CGS (right panel). 
   The errorbars are computed with the lognormal mock method. 
   The dashed line shows the best-fit model from Eq.~\ref{eq:ximodel}. 
   The shaded area represents the $68 \%$ posterior uncertainties provided by
    the MCMC analysis.}
  \label{fig:xiall}
\end{figure*}


\subsubsection{The BAO distance constraint}
\label{subsec:BAO}
The distance constraint is entirely contained in $\alpha$ (Eq.~\ref{eq:ximodel}). 
Correcting for the geometric distortions, it can be derived through the following
approximation:
\begin{equation}
  D_V(\bar{z}) \, = \, \alpha D_V^{fid}(\bar{z}) \left( \frac{r_s}{r_s^{fid}} \right) \Mpc \, .
  \label{eq:Dvalpha}
\end{equation}
Eq.~\ref{eq:Dvalpha} states that the distance constrained at the mean
redshift of the catalogue is $\alpha$ times the distance at the
fiducial cosmology, scaled to the ratio between the true and fiducial
sound horizon.  $D_V(\bar{z})$ is the isotropic volume distance
calculated at the mean redshift of the catalogue:
\begin{equation}
  D_V(z) \, = \, \left[ (1+z)^2 D_A^2(z) \frac{cz}{H(z)}
    \right]^{1/3} \, ,
  \label{eq:Dvdef}
\end{equation}
where $c$ is the speed of light, $D_A(z)$ is the angular diameter
distance and $H(z)$ is the Hubble parameter at redshift $z$.  If we
assume that the true value of the sound horizon is known, we can
directly measure a distance; otherwise we can exploit only an
\emph{uncalibrated} version of the standard ruler technique, measuring
the ratio $D_V(z)/r_s$. In the following, cosmological constraints
will be obtained using both the calibrated and the uncalibrated
distance estimators.


\section{Results}
\label{sec:res}


\subsection{Sample covariance}
Firstly, we test the covariance in the parameter estimation among the
two catalogues with data in common, the Main-GCS and the Lowz-GCS. We
construct $100$ lognormal realizations extracting samples according to
the sample selection function. Then we add the same fraction of data
in common between the samples (almost $2/3$ of the Main-GCS).  After
measuring the 2PCF for each realization, we fitted a simple two-parameter
model for the $\xi(r)$:
\begin{equation}
  \xi(r)=b^2\xi_{DM}(\alpha r) \, ,
  \label{eq:xilognormal}
\end{equation}
where $b$ is the bias factor and $\alpha$ is the shift parameter.  We
then calculate the correlation index $\rho$ as the ratio
$C_{ij}/\sqrt{C_{ii}*C_{jj}}$ with
$C_{ij}=<\alpha_i\alpha_j>-<\alpha_i><\alpha_j>$. We find a moderate
correlation, with $\rho=0.402$.


\subsection{Clustering measurements}
\label{subsec:clmeasure}
\begin{figure*}
  \begin{center}
  \includegraphics[width=\textwidth]{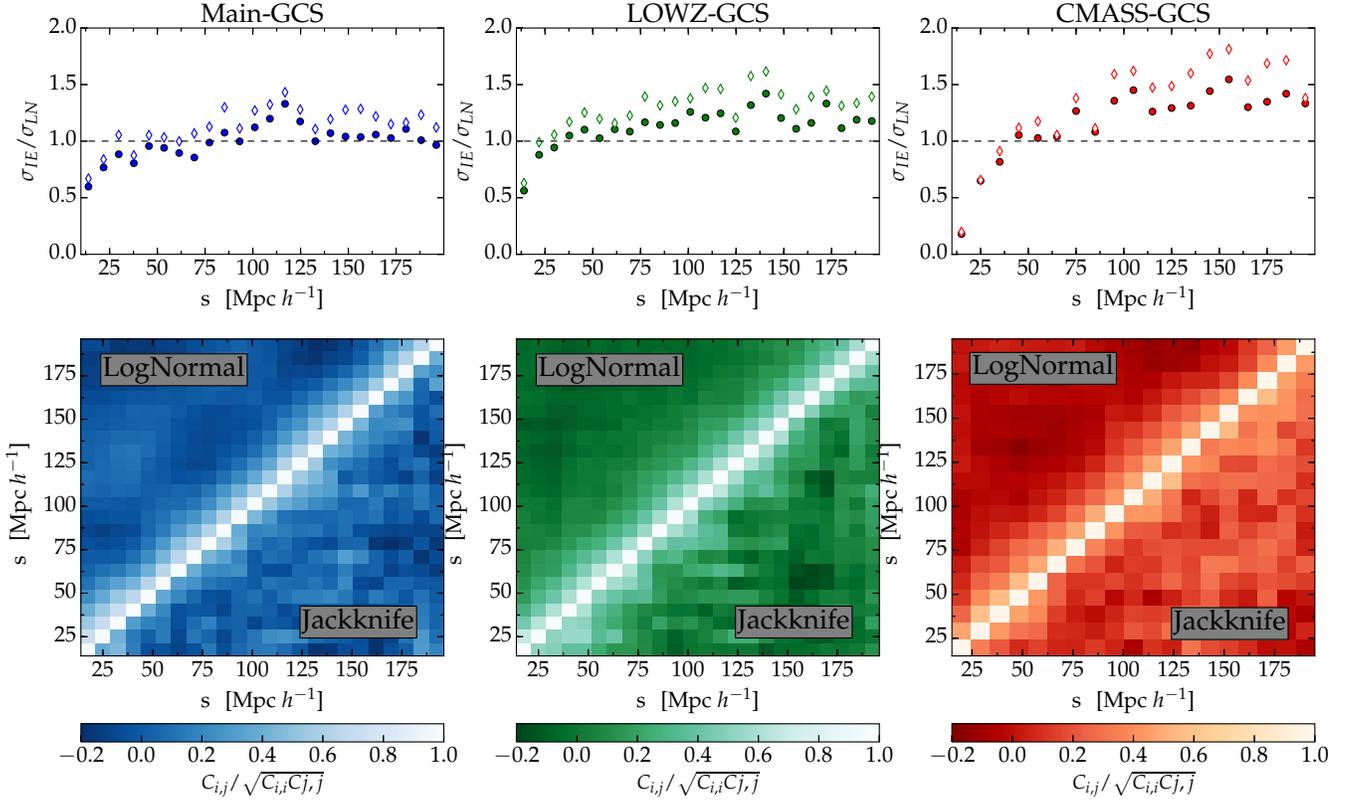}
  \end{center}
  \caption{\textit{Upper panels}: ratio of internal error (jackknife: filled circles,
  bootstrap: open diamonds) and lognormal principal diagonal
    square roots. At the scales of interest for the fit, internal
    error methods provide a conservative error estimate.
    \textit{Lower panels}: correlation matrices ($C_{i,j}/\sqrt{C_{i,i}C_{j,j}}$) from lognormal
    realizations (upper diagonal part) and from jackknife estimates
    (lower diagonal part).  The $1000$ lognormal mocks provide a less
    scattered covariance matrix with respect to the jackknife method.
    This is due mainly to the low number of bins ($50$) used when
    resampling the catalogue.}
  \label{fig:cov}
\end{figure*}

In Fig. \ref{fig:xiall} we present the measured 2PCF for the three
galaxy cluster samples considered.  The errorbars shown are the ones
computed with the lognormal mock method. Measures are robust when
changing the modelization of the radial selection function.  The
clustering signal is well determined despite the sparseness of the
samples.  The measured 2PCFs are all consistent with each other in
terms of the BAO peak position, though a significant difference in the
bias is measured (see Table \ref{tab:samples}).  As already pointed
out in \S~\ref{subsec:covmat}, we consider different methods to
compute the covariance matrix.  In the case of internal errors, we
divide each cluster catalogue in $100$ samples to get the jackknife
estimate, and re-sample them $200$ times to exploit the bootstrap
technique.  We use instead $1000$ lognormal mock realizations.
Results are shown in Fig.~\ref{fig:cov}. In the top panels we compare
internal estimates (filled circles for jackknife, open diamonds for
bootstrap) and lognormal mock estimates of the square root of the
principal diagonal values of the covariance matrix. Internal methods
provide conservative estimates of the errors, that become less biased
for larger number of objects. Both the jackknife and bootstrap
estimates are robust when changing the number of subsamples, or
resamplings. Nevertheless, a large number of realizations makes the
covariance matrix less scattered. This can be seen in the bottom
panels of Fig.~\ref{fig:cov}: the covariance matrix obtained with
lognormal mocks is smoother compared to the jackknife one. In the
following we adopt the lognormal mock estimate of the covariance
matrix as the reference.  Covariance matrices from internal estimators
will be used to check for consistencies in the parameter determination
and BAO peak detection.

\begin{figure*}
  \begin{center}
    \includegraphics[width=\textwidth]{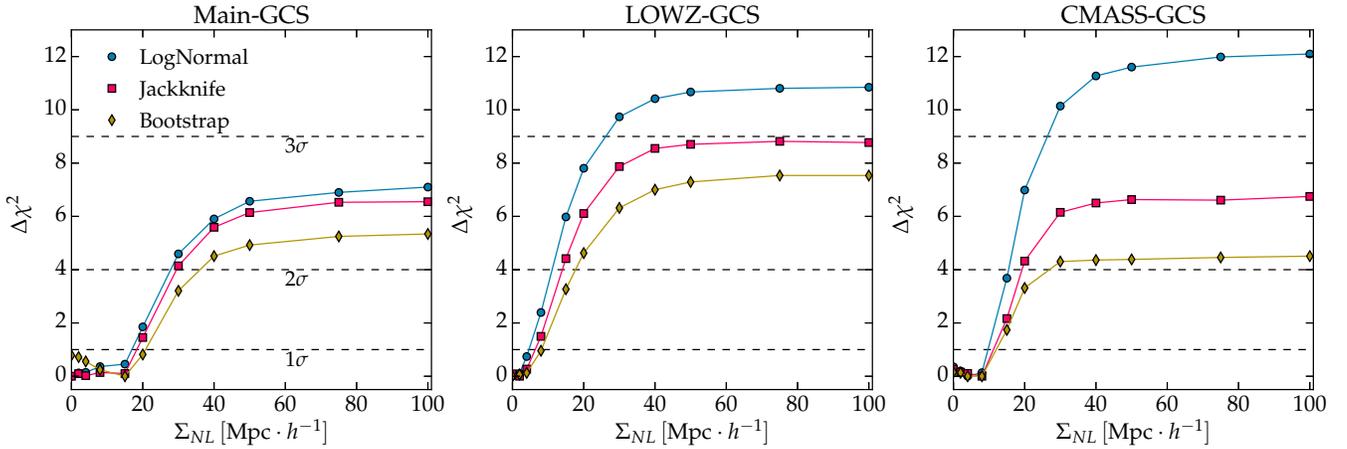}
  \end{center}
  \caption{The values of $\Delta \chi^2$ as a function of
    $\Sigma_{NL}$ for the three cluster samples -- Main-GCS (left
    panel), LOWZ-GCS (centeral panel), and CMASS-GCS (right panel) --
    and for the three covariance matrix definitions -- lognormal mocks
    (blue filled circles), jackknife (red squares) and bootstrap
    (yellow diamonds). The values span from a linear power spectrum
    ($\Sigma_{NL} \, = 0 \, \Mpch$) to a power spectrum model with no
    BAO ($\Sigma_{NL} \, \rightarrow \infty $). Detections of the BAO
    are well over $2\sigma$ in all cases.}
  \label{fig:deltachi2}
  
\end{figure*}
\begin{table}
  \centering
    \caption{Values of the shift parameter $\alpha$ for the three
    catalogues, obtained using covariance matrices from lognormal
    mocks, jackknife and bootstrap subsampling, from column $2$ to
    column $4$. All the results have been obtained by fitting the 
    2PCF from $40 \Mpch$ to $200 \Mpch$, and fixing $\Sigma_{NL}=4\Mpch$.}
  \begin{tabular}{l|c|c|c} 
    \hline
    \hline
    Sample Name & LogNormal & Jackknife & Bootstrap \\
    \hline
    Main-GCS  & $0.97 \pm 0.06$ & $0.97 \pm 0.08$ & $0.98 \pm 0.08$ \\
    LOWZ-GCS  & $0.99 \pm 0.03$ & $0.99 \pm 0.04$ & $0.99 \pm 0.05$ \\
    CMASS-GCS & $0.99 \pm 0.03$ & $0.99 \pm 0.06$ & $0.99 \pm 0.08$  \\
    \hline
    \hline
  \end{tabular}
  \label{tab:alphares}
\end{table}


\subsection{Distance constraints}
\label{subsec:distconstr}
We get distance constraints by fitting the BAO peak position in the
measured 2PCF at different redshifts.  The BAO feature is clearly
detected for all samples. Results of the fits using the different
definitions of the covariance matrix are reported in Table
\ref{tab:alphares}.  The $\alpha$ values estimated in the three
redshift bins are all consistent. The precision in the detection degrades
when using jackknife or bootstrap covariance matrices, as expected.
The model depends on the parameter $\Sigma_{NL}$, that describes the
degradation of the BAO feature in the power spectrum.  We fit the
measured 2PCF by changing this parameter from $0$ to $100 \Mpch$.
Fig.~\ref{fig:deltachi2} shows the values of $\Delta\chi^2$ as a
function of $\Sigma_{NL}$. Each point represents the 
difference between the minimum $\chi^2$ at each
$\Sigma_{NL}$ and the absolute minimum of the curve.  We do this for
all the three samples and for the three error definitions. This allows
us to determine the significance of our detection. Indeed, for high
values of $\Sigma_{NL}$, the BAO peak completely disappears
($\xi(r)\rightarrow \xi_{nw}$, see Eq.~\ref{eq:pkmodel}).  The
significance of the BAO detections result to be well above $2 \sigma$
for all the considered samples.  We cannot distinguish between models
with $\Sigma_{NL} < 8 \Mpch$, that are all consistent within
$1\sigma$. Nevertheless, Fig.~\ref{fig:deltachi2} clearly indicates
that galaxy clusters have a more limited non-linear contribution at
the BAO scales with respect of other tracers, such as galaxies (the
$\Delta\chi^2$ minima are in some cases at $\Sigma_{NL}=0\Mpch$). This
result confirms what found in \citet{paper:veropalumbo2014}. A
practical consequence is that the density field reconstruction seems
not crucial in the BAO distance constraints from galaxy clusters. We
will return to this aspect in \S~\ref{subsec:comparison}, while a more
detailed investigation is postponed to a forthcoming work. Hereafter we
consider the lognormal results with $\Sigma_{NL}=4 \Mpch$ as our
reference distance constraint.

\begin{figure}
  \begin{center}
    \includegraphics[width=\columnwidth]{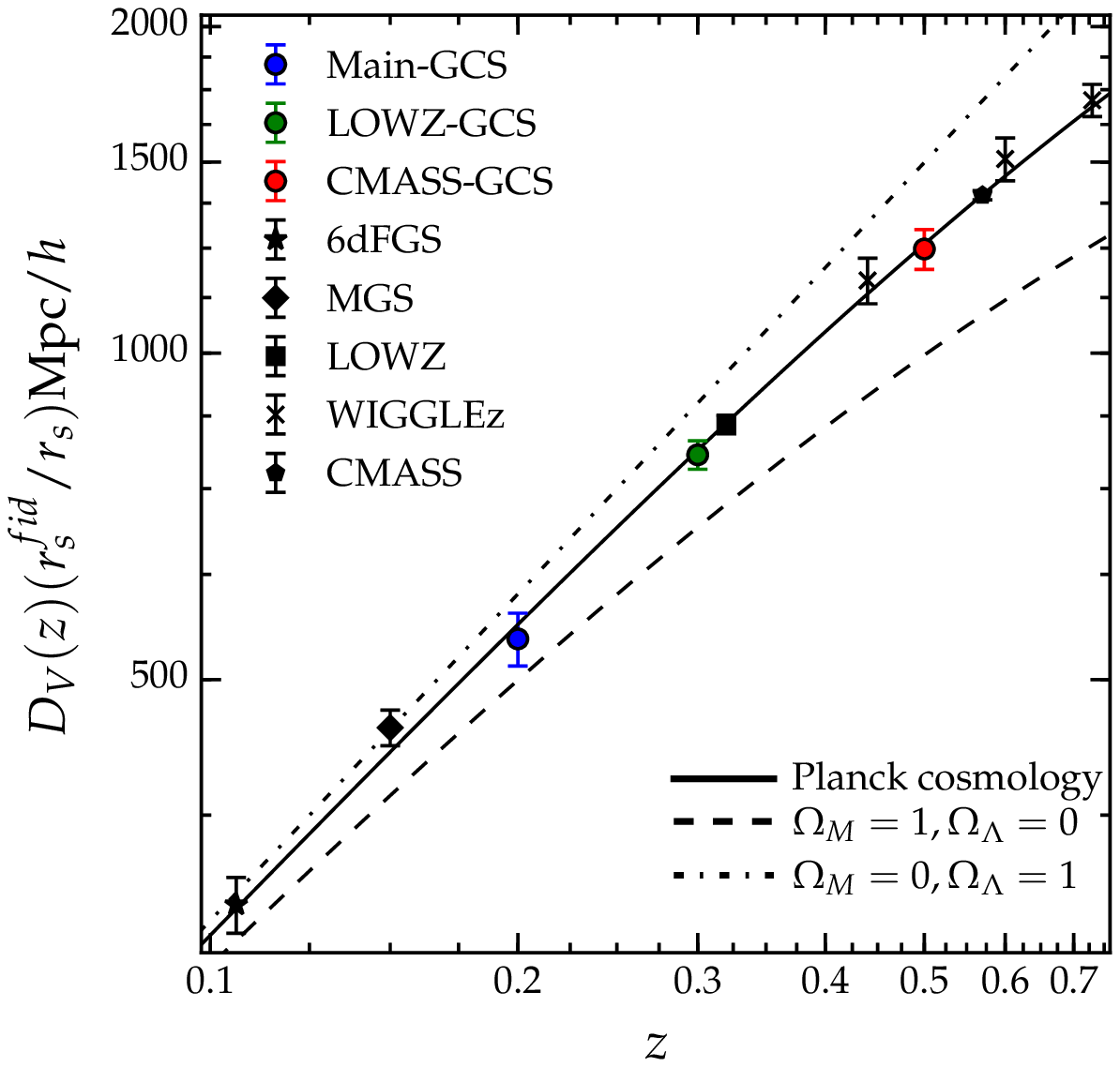}
  \end{center}
  \caption{The distance $D_V(z)/(r_s^{fid}/r_s)$ -- redshift relation.
    The coloured points show our measurements at redshifts $0.2, 0.3,
   0.5$. Other black symbols correspond to other distance constraints from
    galaxy surveys: 6dFGS \citep[star]{paper:beutler2011}, MGS
    \citep[diamond]{paper:ross2015} , BOSS LOWZ and CMASS
    \citep[square and pentagon, respectively]{paper:anderson2014}
    and WiggleZ \citep[crosses]{paper:kazin2014}. The black curve is the $D_V(z)$
    prediction for the $\Lambda$CDM cosmology with the Planck parameters
    \citep{paper:planck2013}. We also show distance prediction 
    in a flat, matter-only Einstein-De Sitter universe (black dashed curve) 
    and for a De Sitter universe with $\Omega_{\Lambda}=1$ (black dot-dashed curve)
 }
  \label{fig:Dv}
\end{figure}

We measure the following values: $D_V(z=0.2)(r_s^{fid}/r_s) = 545 \pm 31 \Mpch $, 
    $D_V(z=0.3)(r_s^{fid}/r_s) = 806 \pm 24 \Mpch $ and 
    $D_V(z=0.5)(r_s^{fid}/r_s) = 1247 \pm 53 \Mpch $.
Fig.~\ref{fig:Dv} shows the distance-redshift
diagram. The coloured points are the isotropic distance estimates for
the Main-GCS (blue), the LOWZ-GCS (green) and the CMASS-GCS (red)
sample, respectively. The other black symbols show some $D_V$
estimates for galaxy samples from literature:
6dFGS survey \citep[black star]{paper:beutler2011},
Main galaxy sample (MGS) from SDSS DR7
\citep[black diamond]{paper:ross2015}, BOSS LOWZ and CMASS 
\citep[black square and pentagon, respectively]{paper:anderson2014} and WiggleZ 
\citep[black cross]{paper:kazin2014}.  
The black curve is the theoretical prediction for the Planck 2013 $\Lambda$CDM 
cosmology \citep{paper:planck2013}. 
As it can be seen, our results are fully consistent with previous
measurements from galaxy surveys, and with standard $\Lambda$CDM
predictions.


\subsection{Cosmological implications}
\label{subsec:cosmoimpl}
We can use the distance measurements described above to derive constraints
on cosmological parameters. Using only the three measures obtained
with our cluster samples we do not expect to get constraints
competitive with the ones obtained by combining larger galaxy samples
with different probes. The aim here is just to check the consistency
of our measurements with the predictions of the standard cosmological
framework.

As reported in \S~\ref{subsec:BAO}, we consider two methods to derive
cosmological constraints, that is the \emph{calibrated} and the
\emph{uncalibrated} distance estimators. In the first case, we use the
Planck value of the sound horizon, $r_s=147.34 \pm 0.65 \Mpc$, to calibrate the
BAO distance measure. We have $D_v(z=0.2)=800 \pm 50
  \Mpc$, $D_V(z=0.3)=1183 \pm 35 \Mpc$ and $D_V(z=0.5) = 1832 \pm 55
  \Mpc$.  In the second approach the sound horizon is a function of
cosmological parameters, through the interpolation formula given by
\citet{paper:anderson2014}.  In this case we get
    $r_s/D_V(z=0.2)=0.18 \pm 0.01$, $r_s/D_V(z=0.3)=0.124 \pm 0.004$
    and $r_s/D_V(z=0.5)=0.080 \pm 0.002$.  The
    value of $\Omega_b$ is kept fixed to the best-fit Planck value
    \citep{paper:planck2013}: $\Omega_b=0.049$.

With both the methods we test some cosmological scenarios. Specifically,
we constrain the cosmological parameters that enter the Hubble
function, $H(z)$.  In fact, the quantity $D_V$ is a function of $H(z)$
and of the angular diameter distance $D_A(z)$, which in turn depends
on the comoving distance $D_C(z) = \int_0^{z} \mathrm{d}z' c / H(z')$
(Eq.~\ref{eq:Dvdef}).  Results of all the fits are summarized in Table
\ref{tab:cosmopars}, Fig.~\ref{fig:contours}, where we show constraints
of cosmological parameters and in Fig.~\ref{fig:dist}, where we 
show the best-fit values of $\alpha$ (Eq.~\ref{eq:Dvalpha}) for the 
two types of distances and for four different cosmological scenarios.


\subsubsection{$\Lambda$CDM models}
\begin{figure*}
  \begin{center}
   \includegraphics[width=0.8\textwidth]{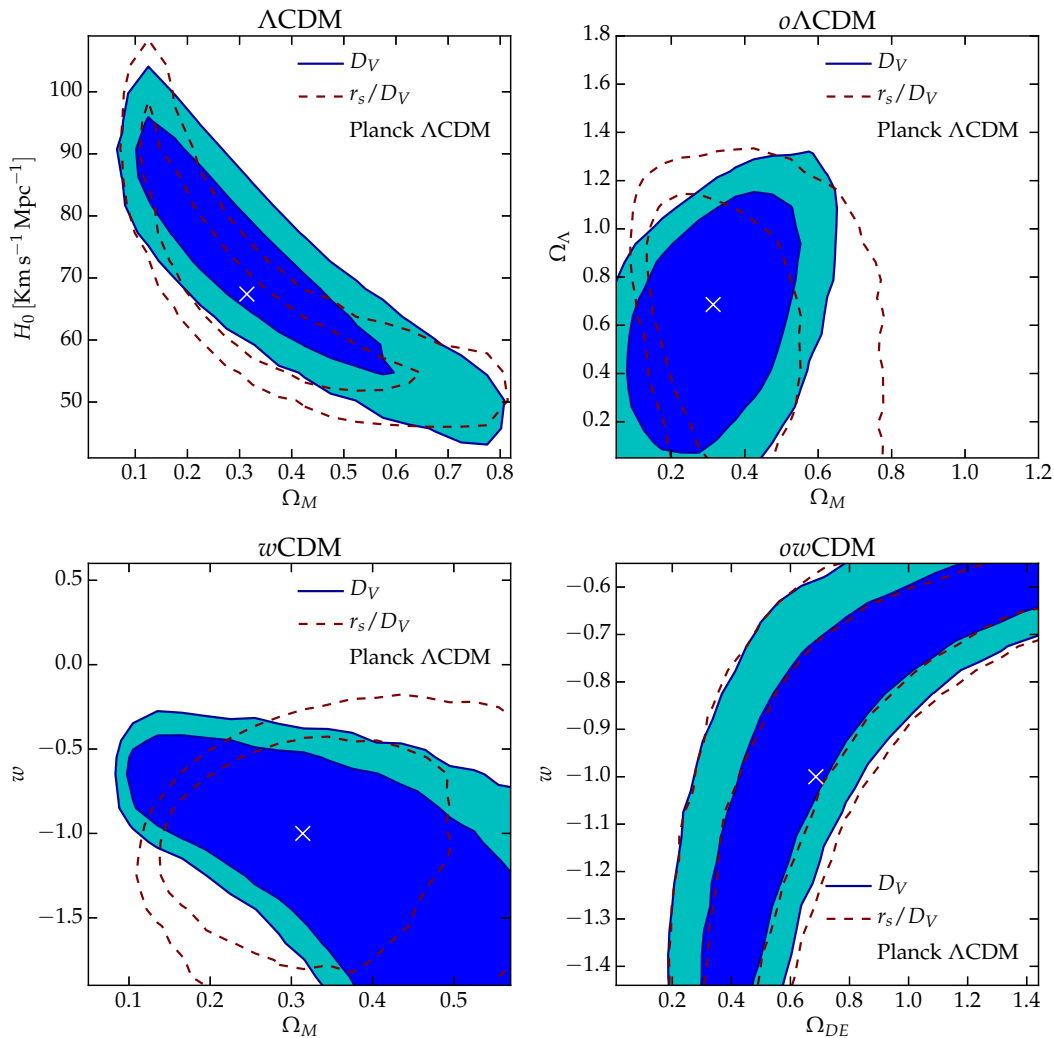}
  \end{center}
  \caption{$1-2\sigma$ confidence contours for parameters assuming different cosmological models:
    $\Omega_M$-$H_0$ plane in the $\Lambda$CDM model (upper left panel), 
    $\Omega_M-\Omega_{\Lambda}$ plane in the $o\Lambda$CDM model (upper right panel),
    $\Omega_M-w$ plane in the $w$CDM  model (lower left panel) and
    $\Omega_{DE}-w$ plane in the $ow$CDM model(lower right panel).
    The red dashed contours show results from the uncalibrated 
    distance $r_s/D_V$, the blue for the calibrated distance $D_V$. 
    The best fit value for Planck $\Lambda$CDM cosmology
    is also reported (black cross).}
  \label{fig:contours}
\end{figure*}

\begin{figure*}
  \begin{center}
     \includegraphics[width=0.8\hsize]{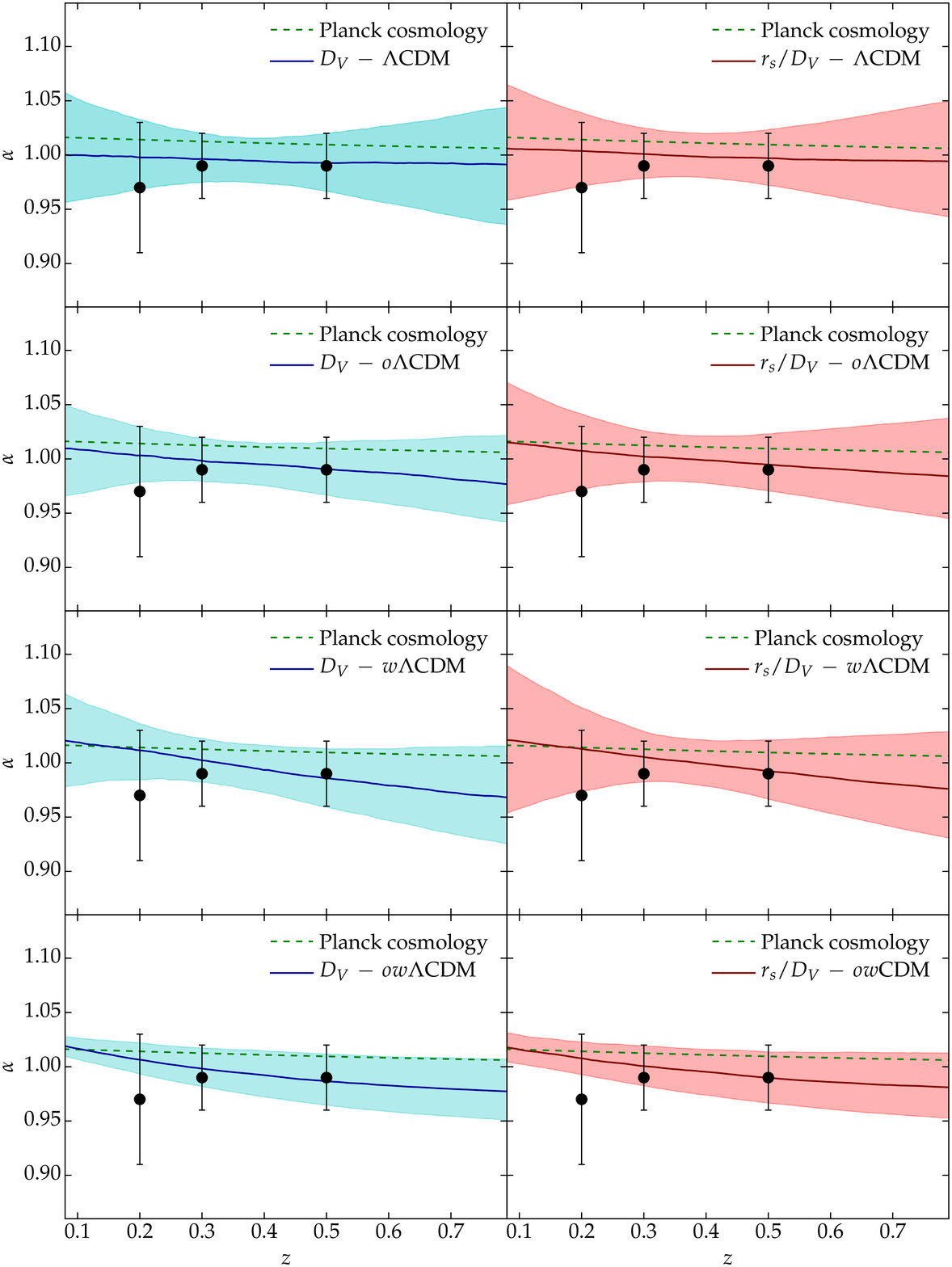}

  \end{center}
  \caption{The best-fit values of $\alpha(z) \equiv D_V(z)/r_s/
    (D_V(z)/r_s)^{fid}$ (solid curves) and $68\%$ model error (shaded
    areas) using the calibrated and uncalibrated BAO distances (black dots,
    respectively in the left and right panels).
    We show results for the four cosmological model tested:
    $\Lambda$CDM (first row), $o\Lambda$CDM (second row),
    $w$CDM (third row) and $ow$CDM (fourth row).
    The Planck best-fit values
    for $\alpha$ are shown by the green dashed curves.  For the fit with
    calibrated distances, we assumed the Planck value for the sound
    horizon.}
 \label{fig:dist}
\end{figure*}

The simplest model we test is the flat $\Lambda$CDM Universe (equation
of state $w = -1$), with a negligible contribution of radiation. 
In this case the Hubble function reads:
\begin{equation}
  H^2(z) / H^2_0 = \Omega_M (1+z)^{3} + \Omega_{\Lambda}.
  \label{eq:lcdm}
\end{equation} 
Since the curvature $\Omega_k =
1-\Omega_M-\Omega_{\Lambda}-\Omega_{r}$ is fixed to zero,
$\Omega_{\Lambda}$ is a function of $\Omega_M$.  We fit our distance
constraints against the pair $\lbrace \Omega_M , H_0 \rbrace$.  We
impose a large uniform prior on both parameters: $\mathcal{U}(0,1)$ for $\Omega_M$
and $\mathcal{U}(30,120)$ for $H_0$.
Here $\mathcal{U}(a,b)$ is the uniform distribution, equal to $0$ outside ranges $a,b$.
We find $\Omega_M =0.32^{+0.21}_{-0.14} $ and $H_0 = 72^{+13}_{-13} \, \Hunit$ for the
calibrated distance indicator, and $\Omega_M =0.32^{+0.22}_{-0.15} $
and $H_0 = 64^{+17}_{-9} \, \Hunit$ for the uncalibrated case.
Results are summarized in Fig.~\ref{fig:contours} (upper left panel), 
where we show the $1-2\sigma$ confidence contours for the parameters $\Omega_M-H_0$, 
and in the upper row in Fig.~\ref{fig:dist}.  
The Planck cosmology is well compatible with our results, in both cases. 
Our constraints are broad, due to our distance uncertainties 
and to our limits in redshift. As we will show in particular 
for the next cases, high-redshift distance measures
significantly help in measuring the geometry of the Universe.

\subsubsection{$o\Lambda$CDM models}
Here we test our measurements against a non-flat Universe with CDM and
cosmological constant. In this case the Hubble equation becomes:
\begin{equation}
  H^2(z) / H^2_0 = \Omega_M (1+z)^{3} +
  \Omega_{\Lambda}+\Omega_k(1+z)^{2}.
  \label{eq:olcdm}
\end{equation}

We fit the combination of parameters $\lbrace \Omega_M ,
\Omega_{\Lambda}, H_0 \rbrace$, using a flat prior on $\Omega_M$, 
$\Omega_{\Lambda}$ and a broad Gaussian prior on $H_0$ $\mathcal{N}(\mu,\sigma)$
with mean $\mu$ centered on the Planck value $67 \Hunit$, and
standard deviation $\sigma$ of $ 20 \Hunit$.
See Tab.~\ref{tab:cosmopars} for more information on the adopted priors.
We find $\Omega_k = 0.05 \pm 0.40$ for the $D_V$ fit and $\Omega_k =
-0.01_{-0.36}^{+0.34}$ for the fit using $r_s/D_V$. In
Fig.~\ref{fig:contours} (upper right panel) 
we show the $1-2 \sigma$ confidence contours for
the combination $\Omega_M-\Omega_{\Lambda}$, marginalized over $H_0$.
We confirm at $1 \sigma$ the necessity of a negative pressure
component in the cosmological model.  The different degeneracy
directions obtained with the two methods are due to the introduction
of the cosmological dependence of the sound horizon, that depends only on
$\Omega_M h^2$ and $\Omega_b h^2$. In Fig.~\ref{fig:dist}, second row,
we can appreciate the increasing difference between the Planck
$\Lambda$CDM best-fit model (dashed green curve) and the best-fit model
obtained with our data (blue solid line), going to high redshifts. The difference is
well inside $1 \sigma$, due to our distance uncertainties, though a
little tension with the Planck results can be noticed.

\begin{table*}
  \centering
    \caption{Summary of the cosmological parameters obtained by fitting the calibrated BAO
    distance $D_V$ and the uncalibrated BAO distance $d_z$. We report priors used in the fitting
    procedures.}
    \begin{tabular}{l|c|c|c|c|c|c|c|c|c|} 
    \hline
    \hline
     & Distance & \multicolumn{2}{c}{$H_0 [\Hunit]$} & \multicolumn{2}{c}{$\Omega_M$} & 
    \multicolumn{2}{c}{$\Omega_{\Lambda} / \Omega_{DE}$} & \multicolumn{2}{c}{$w$} \\
    \hline
    \hline
     & & Prior & Posterior & Prior & Posterior & Prior & Posterior & Prior & Posterior \\
     \hline
    \multirow{2}{*}{$\Lambda$CDM} & $D_V$ & $\mathcal{U}(30,120)$ & $72^{+13}_{-13}$ & $\mathcal{U}(0,1)$ &$0.32^{+0.21}_{-0.14}$ & $-$ & $-$ & $-$ & $-$\\
  								  & $d_z$ & $"$ & $64^{+17}_{-9}$ & $"$ & $0.32^{+0.22}_{-0.15}$ & $-$ & $-$ & $-$ & $-$ \\
    \hline
    \multirow{2}{*}{$o\Lambda$CDM} & $D_V$ & $\mathcal{N}(67,20)$ & $70^{+11}_{-10}$ & $\mathcal{U}(0,1)$ & $0.32^{+0.15}_{-0.14}$ & $\mathcal{U}(0,1.5)$ & $0.63^{+0.30}_{-0.29}$ &$-$ & $-$\\
						    & $d_z$ & $"$ & $62^{+14}_{-12}$ & $"$ & $0.36^{+0.19}_{-0.13}$ & $"$ & $0.65^{+0.33}_{-0.37}$ &$-$ & $-$ \\
   \hline
   \multirow{2}{*}{$w$CDM} & $D_V$ & $\mathcal{N}(67,20)$ & $69^{+11}_{-10}$ & $\mathcal{U}(0,1)$ & $0.39^{+0.16}_{-0.18}$ & $-$ & $-$ & $\mathcal{U}(-2,0)$ & $-1.15^{+0.43}_{-0.54}$  \\
						    & $d_z$ & $"$ & $64^{+12}_{-8}$ & $"$ & $0.38^{+0.19}_{-0.12}$ & $-$ & $-$ & $"$ & $-1.01^{+0.44}_{-0.44}$  \\
    \hline
   \multirow{2}{*}{$o w$CDM} & $D_V$ & $\mathcal{N}(67,2)$ & $67.0^{+1.4}_{-1.4}$ & $\mathcal{N}(0.31,0.02)$ & $0.31^{+0.01}_{-0.01}$ & $\mathcal{U}(0,1.5)$ & $0.66^{+0.45}_{-0.24}$ & $\mathcal{U}(-1.5,0)$ &$-0.91^{+0.23}_{-0.38}$ \\
							& $d_z$ & $"$ & $67.0^{+1.4}_{-1.4}$ & $"$ & $0.31^{+0.01}_{-0.01}$ & $"$ & $0.66^{+0.46}_{-0.21}$ & $"$ & $-0.88^{+0.24}_{-0.36}$ \\
    \hline
    \hline
    \end{tabular}
  \label{tab:cosmopars}
\end{table*}


\subsubsection{$w$CDM models}
We now test our data against a flat Universe with CDM and dark energy,
with dark energy density changing with time. We parametrize the dark energy
density time dependence with a constant $w$ equation of state.
The Hubble equation in this case is:
\begin{equation}
  H^2(z) / H^2_0 = \Omega_M (1+z)^{3} + \Omega_{DE}(1+z)^{3(1+w)}.
  \label{eq:wcdm}
\end{equation} 
This model turns into standard $\Lambda$CDM cosmology imposing $w=-1$.
We fit the combination of parameters $\lbrace \Omega_M , w, H_0
\rbrace$, since, as in the $\Lambda$CDM case, the value of
$\Omega_{DE}$ at the present time is fixed by the relation
$\Omega_{DE} = 1- \Omega_M$.  We assume a flat prior on $\Omega_M$ and
$w$ and a broad Gaussian prior on $H_0$, centered at the Planck
value $67 \Hunit$, with standard deviation of $ 20 \Hunit$.  We find
$w = -1.15_{-0.54}^{+0.43}$ for the calibrated distance fit, and
$w = -1.01_{-0.44}^{+0.44}$ for the uncalibrated one. Confidence
contours up to $2\sigma$ for the pair $\Omega_M-w$ are shown in
Fig.~\ref{fig:contours} (lower left panel), while the best-fit
corresponding value of $\alpha$ is reported in the third row
of Fig.~\ref{fig:contours}.


\subsubsection{$ow$CDM models}
The most general case we consider is the one with a non-flat Universe
with time-dependent dark energy density:
\begin{equation}
  H^2(z) / H^2_0 = \Omega_M (1+z)^{3} + \Omega_{DE}(1+z)^{3(1+w)}+\Omega_k(1+z)^2.
  \label{eq:owcdm}
\end{equation}
We vary the parameters $\lbrace \Omega_M , \Omega_{DE}, w, H_0
\rbrace$, keeping totally free $\Omega_{DE}$ and $w$, and assuming a
Gaussian prior for $\Omega_M$, centered on $0.31$ with a
standard deviation of $ 0.02$, and for $H_0$ centered on $67 \Hunit$ with a
small standard deviation of $ 2 \Hunit$, respectively.  In the case of
the uncalibrated distance measure, the strong priors on $\Omega_M$ and
$H_0$ resume in an almost constant value for the sound horizon. Since
the central values for these two parameters are the ones from Planck,
the value of the sound horizon is centered on the Planck value
too. This explains the similarity in the bottom row of
Fig.~\ref{fig:dist}, comparing the calibrated and uncalibrated version
of the fit.  Fig.~\ref{fig:contours} (lower right panel) 
shows the degeneracy between the
parameters $\Omega_{DE}$ and $w$. Even with the assumption of 
these strong priors, no clear constraint can be
extracted in this case. In Table \ref{tab:cosmopars} we report the
$16-84 th$ percentile interval of the parameter posterior
distribution.


\subsection{Comparison with previous measurements}
\label{subsec:comparison}

\begin{figure}
  \begin{center}
       \includegraphics[width=\columnwidth]{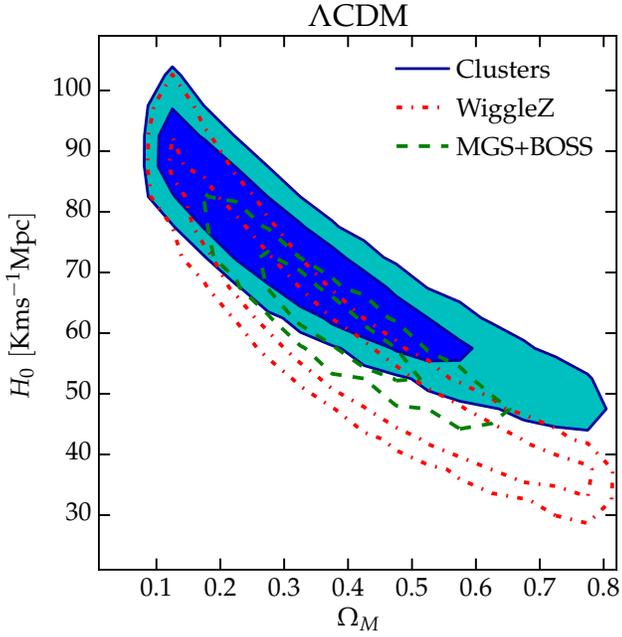}
  \end{center}
  \caption{Comparison of the $1-2 \sigma$ confidence contours in the
    $\Omega_M-H_0$ plane between our work (blue filled contours) and previous
    measurements from galaxy samples -- WiggleZ (red dot-dashed contours) and MGS+BOSS
    (green dashed contours).}
  \label{fig:LCDMcomparison}
\end{figure}

Fig.~\ref{fig:LCDMcomparison} shows the posterior $1-2\sigma$
confidence contours of $\Omega_M$--$H_0$ parameters, obtained with the
calibrated distance estimators (see \S~\ref{subsec:BAO}) from the BAO
of the galaxy cluster samples considered in this work, and of the
galaxy samples of MGS+BOSS and WiggleZ.  As one can see, our
constraints are consistent with previous estimates. Our uncertainties
appear slightly better than the ones obtained by modelling the
post-reconstruction WiggleZ clustering \citep{paper:kazin2014},
despite the paucity of our samples, while they are broader with
respect to the ones from the BOSS survey. This is expected, since our
BCGs represent a subsample of the BOSS galaxy survey. Indeed, the
measurements obtained with these BOSS galaxy catalogues provide the
best BAO distance constraints to date, both in term of accuracy and of
BAO reconstruction and modelling techniques\footnote{The 2PCF for the
  galaxies of the CMASS sample pre- and post-reconstruction, together
  with covariance matrices, are publicly available
  \url{https://www.sdss3.org/science/boss_publications.php}.}.

Compared to clusters, we find a lower value of the bias factor for the
BOSS galaxies.  This result is expected and implies that our cluster
catalogues cannot be simply considered as random subsamples of the
galaxy catalogue, but they identify the highest peaks of the density
field, confirming our previous results \citep{paper:veropalumbo2014}.

\begin{figure}
  \begin{center}
       \includegraphics[width=\columnwidth]{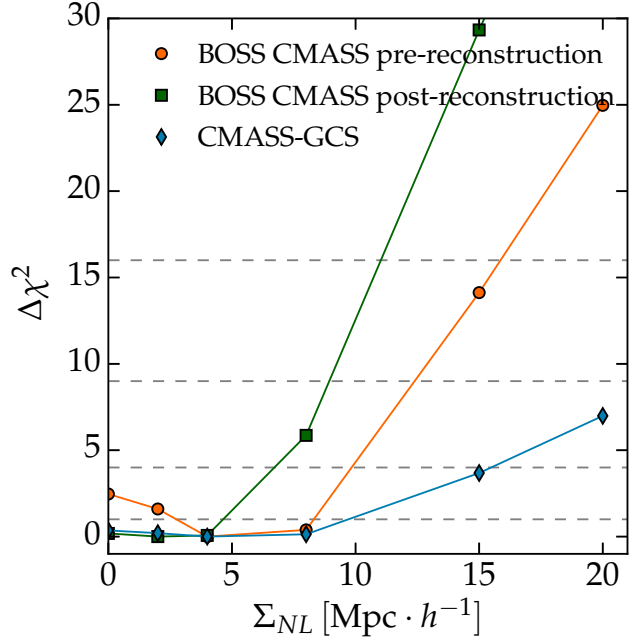}
  \end{center}
  \caption{The same as Fig.~\ref{fig:deltachi2} but for BOSS CMASS
    before (orange filled circles) and after reconstruction (green
    filled squares) \citep{paper:anderson2014} and CMASS-GCS, with
    lognormal mock covariance matrix (blue filled diamonds).  The high
    significance of the BAO detection is evident for galaxies,
    compared to our results.  The reconstruction process helps in
    reducing non-linear effects, shifting the $\Delta\chi^2$ minimum
    to $4 \Mpch$ in the post-reconstruction measurement.}
  \label{fig:galcl}
\end{figure}

Finally, we reanalyze the BOSS CMASS clustering data with the method
described in \S~\ref{subsec:distconstr}.  Fig.~\ref{fig:galcl} shows
the values of $\Delta\chi^2$ as a function of $\Sigma_{NL}$ for the
BOSS CMASS pre- and post-reconstruction samples, and for our CMASS-GCS
sample. Firstly, our estimated value of the BAO detection significance
for the BOSS CMASS data is consistent with the value claimed by
\citet{paper:anderson2014}. Moreover, this analysis highlights the
impact of the density field reconstruction technique, that shifts the
best-fit value of $\Sigma_{NL}$ from $8 \Mpch$ to $4 \Mpch$ and lower,
improving BAO distance constraints. On the other hand,
  galaxy clusters trace a more linear density field with respect to
  galaxies.  Thus, we expect that the BAO constraints would not
improve significantly as a result of the reconstruction technique,
though we cannot draw any robust conclusion about this aspect due to
our measure uncertainties.


\section{Conclusions}
\label{sec:conclusion}
In this work we obtained the first observational constraints on the
distance-redshift relation using only the clustering properties of
galaxy clusters. Specifically, we measured the 2PCF of the largest
spectroscopic galaxy cluster samples to date, in three redshift ranges,
and extract consmological constraints from the position of the BAO
peak.  The catalogues have been constructed by matching the BCGs from
the WHL catalogue \citep{paper:whl2012}, with spectra from SDSS DR7
\citep{paper:sdssdr7} and SDSS DR12 \citep{paper:sdssdr12}.  This
allowed us to construct three catalogues of galaxy clusters --
Main-GCS, LOWZ-GCS and CMASS-GCS -- at the median redshift $z=0.2$,
$0.3$ and $0.5$, respectively.
We estimated the covariance matrix using both internal error estimators 
(jackknife and bootstrap) and with the lognormal mock method. 
These estimators provide fairly consistent errors, with internal errors 
more conservative and scattered.  We choose the lognormal mock covariance 
matrix estimate as reference. The BAO feature is detected with a 
significance larger than $2 \sigma$, for all considered samples.  
The derived cosmological constraints are competitive with respect 
to other estimates from the BAO peak obtained using richer galaxy catalogues, 
such as 6dFGS \citep{paper:beutler2011} and WiggleZ \citep{paper:blake2011}.
As expected, our results are instead not comparable in precision with the
constraints coming from the BOSS survey, of which our BCGs represent a
subsample. For the three samples analysed we get: $\alpha(z=0.2) =
0.96 \pm 0.06$, $\alpha(z=0.3) = 0.99 \pm 0.03$ and $\alpha(z=0.5) =
0.99 \pm 0.03$, respectively.  This translates to the uncalibrated
distance estimates: $r_s/D_V(z=0.2) = 0.18 \pm 0.01$, $r_s/D_V(z=0.3)
= 0.124 \pm 0.004 $, $r_s/D_V(z=0.5) = 0.080 \pm 0.002$.  We use the
sound horizon estimate from \citet{paper:planck2013} to calibrate the
distances, obtaining: $D_V(z=0.2) = 800 \pm 50 \Mpc $, $D_V(z=0.3) =
1183 \pm 35$, $D_V(z=0.5) = 1832 \pm 55$.  We then use both the
uncalibrated and calibrated distance estimates to derive cosmological
constraints. Our results are all consistent with the cosmological
model supported by the Planck results \citep[see][]{paper:planck2013}. 
To disentangle cosmological parameter degeneracies we would need higher 
redshift-distance constraints.

This study clearly demonstrates that galaxy clusters are powerful
tracers of the cosmic density field and can be efficiently exploited
for BAO analyses.  Despite the paucity of cluster
  samples, with respect to generally larger galaxy samples, the higher
  values of cluster bias and the fact that their redshifts are less
  distorted by random motions improve the clustering signal, that
  results almost insensitive to non-linear dynamical distortions.
  This reflects in a sharper BAO peak in the 2PCF, close to the
  prediction of linear theory. To further tighten the cosmological
constraints obtained in this work, we plan to combine these clustering
measurements with estimates of the cluster mass function. These
investigations will be presented in a forthcoming paper. The
techniques presented here will be further exploited in the next future
on the increasingly large collections of data expected from new
experiments like e.g. Euclid \citep{paper:laureijs2011,
  paper:amendola2013,paper:sartoris2015}, eBOSS
\citep{paper:dawson2015} and eRosita
\citep{paper:merloni2012}.  Galaxy cluster samples and
  dedicated spectroscopic follow-up will provide in fact an
  independent tracer of the dark matter density field, with respect to
  the typical emission line galaxies, targets of many future
  experiments.

\section*{Acknowledgments}
We acknowledge the grants ASI n.I/023/12/0 ``Attività relative alla fase B2/C per la missione Euclid'',
MIUR PRIN 2010-2011 ``The dark Universe and the cosmic evolution of baryons: from current surveys to Euclid'' and PRIN INAF 2012 ``The Universe in the box: multiscale simulations of cosmic structure''.

\bibliographystyle{mn2e_fix_Williams}
\bibliography{clusterBAO}

\begin{thebibliography}{45}
\expandafter\ifx\csname natexlab\endcsname\relax\def\natexlab#1{#1}\fi

\bibitem[{{Abazajian} {et~al}\mbox{.}(2009){Abazajian}, {Adelman-McCarthy},
  {Ag{\"u}eros}, {Allam}, {Allende Prieto}, {An}, {Anderson}, {Anderson},
  {Annis}, {Bahcall}, \& et~al.}]{paper:sdssdr7}
{Abazajian} K.~N. {et~al.}, 2009, \apjs, 182, 543

\bibitem[{{Aihara} {et~al}\mbox{.}(2011){Aihara}, {Allende Prieto}, {An},
  {Anderson}, {Aubourg}, {Balbinot}, {Beers}, {Berlind}, {Bickerton},
  {Bizyaev}, {Blanton}, {Bochanski}, {Bolton}, {Bovy}, {Brandt}, {Brinkmann},
  {Brown}, {Brownstein}, {Busca}, {Campbell}, {Carr}, {Chen}, {Chiappini},
  {Comparat}, {Connolly}, {Cortes}, {Croft}, {Cuesta}, {da Costa}, {Davenport},
  {Dawson}, {Dhital}, {Ealet}, {Ebelke}, {Edmondson}, {Eisenstein},
  {Escoffier}, {Esposito}, {Evans}, {Fan}, {Femen{\'{\i}}a Castell{\'a}},
  {Font-Ribera}, {Frinchaboy}, {Ge}, {Gillespie}, {Gilmore}, {Gonz{\'a}lez
  Hern{\'a}ndez}, {Gott}, {Gould}, {Grebel}, {Gunn}, {Hamilton}, {Harding},
  {Harris}, {Hawley}, {Hearty}, {Ho}, {Hogg}, {Holtzman}, {Honscheid}, {Inada},
  {Ivans}, {Jiang}, {Johnson}, {Jordan}, {Jordan}, {Kazin}, {Kirkby}, {Klaene},
  {Knapp}, {Kneib}, {Kochanek}, {Koesterke}, {Kollmeier}, {Kron}, {Lampeitl},
  {Lang}, {Le Goff}, {Lee}, {Lin}, {Long}, {Loomis}, {Lucatello}, {Lundgren},
  {Lupton}, {Ma}, {MacDonald}, {Mahadevan}, {Maia}, {Makler}, {Malanushenko},
  {Malanushenko}, {Mandelbaum}, {Maraston}, {Margala}, {Masters}, {McBride},
  {McGehee}, {McGreer}, {M{\'e}nard}, {Miralda-Escud{\'e}}, {Morrison},
  {Mullally}, {Muna}, {Munn}, {Murayama}, {Myers}, {Naugle}, {Neto}, {Nguyen},
  {Nichol}, {O'Connell}, {Ogando}, {Olmstead}, {Oravetz}, {Padmanabhan},
  {Palanque-Delabrouille}, {Pan}, {Pandey}, {P{\^a}ris}, {Percival},
  {Petitjean}, {Pfaffenberger}, {Pforr}, {Phleps}, {Pichon}, {Pieri}, {Prada},
  {Price-Whelan}, {Raddick}, {Ramos}, {Reyl{\'e}}, {Rich}, {Richards}, {Rix},
  {Robin}, {Rocha-Pinto}, {Rockosi}, {Roe}, {Rollinde}, {Ross}, {Ross},
  {Rossetto}, {S{\'a}nchez}, {Sayres}, {Schlegel}, {Schlesinger}, {Schmidt},
  {Schneider}, {Sheldon}, {Shu}, {Simmerer}, {Simmons}, {Sivarani}, {Snedden},
  {Sobeck}, {Steinmetz}, {Strauss}, {Szalay}, {Tanaka}, {Thakar}, {Thomas},
  {Tinker}, {Tofflemire}, {Tojeiro}, {Tremonti}, {Vandenberg}, {Vargas
  Maga{\~n}a}, {Verde}, {Vogt}, {Wake}, {Wang}, {Weaver}, {Weinberg}, {White},
  {White}, {Yanny}, {Yasuda}, {Yeche}, \& {Zehavi}}]{paper:sdssdr8}
{Aihara} H. {et~al.}, 2011, \apjs, 193, 29

\bibitem[{{Alam} {et~al}\mbox{.}(2015){Alam}, {Albareti}, {Allende Prieto},
  {Anders}, {Anderson}, {Anderton}, {Andrews}, {Armengaud}, {Aubourg},
  {Bailey}, \& et~al.}]{paper:sdssdr12}
{Alam} S. {et~al.}, 2015, \apjs, 219, 12

\bibitem[{{Allen}, {Evrard} \& {Mantz}(2011){Allen}, {Evrard}, \&
  {Mantz}}]{paper:allen2011}
{Allen} S.~W., {Evrard} A.~E., {Mantz} A.~B., 2011, \araa, 49, 409

\bibitem[{{Amendola} {et~al}\mbox{.}(2013){Amendola}
  {et~al.}}]{paper:amendola2013}
{Amendola} L., {et~al.}, 2013, Living Reviews in Relativity, 16, 6

\bibitem[{{Anderson} {et~al}\mbox{.}(2014){Anderson}, {Aubourg}, {Bailey},
  {Beutler}, {Bhardwaj}, {Blanton}, {Bolton}, {Brinkmann}, {Brownstein},
  {Burden}, {Chuang}, {Cuesta}, {Dawson}, {Eisenstein}, {Escoffier}, {Gunn},
  {Guo}, {Ho}, {Honscheid}, {Howlett}, {Kirkby}, {Lupton}, {Manera},
  {Maraston}, {McBride}, {Mena}, {Montesano}, {Nichol}, {Nuza}, {Olmstead},
  {Padmanabhan}, {Palanque-Delabrouille}, {Parejko}, {Percival}, {Petitjean},
  {Prada}, {Price-Whelan}, {Reid}, {Roe}, {Ross}, {Ross}, {Sabiu}, {Saito},
  {Samushia}, {S{\'a}nchez}, {Schlegel}, {Schneider}, {Scoccola}, {Seo},
  {Skibba}, {Strauss}, {Swanson}, {Thomas}, {Tinker}, {Tojeiro}, {Maga{\~n}a},
  {Verde}, {Wake}, {Weaver}, {Weinberg}, {White}, {Xu}, {Y{\`e}che}, {Zehavi},
  \& {Zhao}}]{paper:anderson2014}
{Anderson} L. {et~al.}, 2014, \mnras, 441, 24

\bibitem[{{Anderson} {et~al}\mbox{.}(2012){Anderson}, {Aubourg}, {Bailey},
  {Bizyaev}, {Blanton}, {Bolton}, {Brinkmann}, {Brownstein}, {Burden},
  {Cuesta}, {da Costa}, {Dawson}, {de Putter}, {Eisenstein}, {Gunn}, {Guo},
  {Hamilton}, {Harding}, {Ho}, {Honscheid}, {Kazin}, {Kirkby}, {Kneib},
  {Labatie}, {Loomis}, {Lupton}, {Malanushenko}, {Malanushenko}, {Mandelbaum},
  {Manera}, {Maraston}, {McBride}, {Mehta}, {Mena}, {Montesano}, {Muna},
  {Nichol}, {Nuza}, {Olmstead}, {Oravetz}, {Padmanabhan},
  {Palanque-Delabrouille}, {Pan}, {Parejko}, {P{\^a}ris}, {Percival},
  {Petitjean}, {Prada}, {Reid}, {Roe}, {Ross}, {Ross}, {Samushia},
  {S{\'a}nchez}, {Schlegel}, {Schneider}, {Sc{\'o}ccola}, {Seo}, {Sheldon},
  {Simmons}, {Skibba}, {Strauss}, {Swanson}, {Thomas}, {Tinker}, {Tojeiro},
  {Maga{\~n}a}, {Verde}, {Wagner}, {Wake}, {Weaver}, {Weinberg}, {White}, {Xu},
  {Y{\`e}che}, {Zehavi}, \& {Zhao}}]{paper:anderson2012}
{Anderson} L. {et~al.}, 2012, \mnras, 427, 3435

\bibitem[{{Berlind} {et~al}\mbox{.}(2006){Berlind}, {Frieman}, {Weinberg},
  {Blanton}, {Warren}, {Abazajian}, {Scranton}, {Hogg}, {Scoccimarro},
  {Bahcall}, {Brinkmann}, {Gott}, {Kleinman}, {Krzesinski}, {Lee}, {Miller},
  {Nitta}, {Schneider}, {Tucker}, {Zehavi}, \& {SDSS
  Collaboration}}]{paper:berlind2006}
{Berlind} A.~A. {et~al.}, 2006, \apjs, 167, 1

\bibitem[{{Beutler} {et~al}\mbox{.}(2011){Beutler}, {Blake}, {Colless},
  {Jones}, {Staveley-Smith}, {Campbell}, {Parker}, {Saunders}, \&
  {Watson}}]{paper:beutler2011}
{Beutler} F. {et~al.}, 2011, \mnras, 416, 3017

\bibitem[{{Blake} {et~al}\mbox{.}(2011){Blake}, {Davis}, {Poole}, {Parkinson},
  {Brough}, {Colless}, {Contreras}, {Couch}, {Croom}, {Drinkwater}, {Forster},
  {Gilbank}, {Gladders}, {Glazebrook}, {Jelliffe}, {Jurek}, {Li}, {Madore},
  {Martin}, {Pimbblet}, {Pracy}, {Sharp}, {Wisnioski}, {Woods}, {Wyder}, \&
  {Yee}}]{paper:blake2011}
{Blake} C. {et~al.}, 2011, \mnras, 415, 2892

\bibitem[{{Blanton} {et~al}\mbox{.}(2003){Blanton}, {Hogg}, {Bahcall},
  {Brinkmann}, {Britton}, {Connolly}, {Csabai}, {Fukugita}, {Loveday},
  {Meiksin}, {Munn}, {Nichol}, {Okamura}, {Quinn}, {Schneider}, {Shimasaku},
  {Strauss}, {Tegmark}, {Vogeley}, \& {Weinberg}}]{paper:blanton2003}
{Blanton} M.~R. {et~al.}, 2003, \apj, 592, 819

\bibitem[{{Blanton} {et~al}\mbox{.}(2005){Blanton}, {Schlegel}, {Strauss},
  {Brinkmann}, {Finkbeiner}, {Fukugita}, {Gunn}, {Hogg}, {Ivezi{\'c}}, {Knapp},
  {Lupton}, {Munn}, {Schneider}, {Tegmark}, \& {Zehavi}}]{paper:blanton2005}
{Blanton} M.~R. {et~al.}, 2005, \aj, 129, 2562

\bibitem[{{Chuang} {et~al}\mbox{.}(2014){Chuang}, {Zhao}, {Prada}, {Munari},
  {Avila}, {Izard}, {Kitaura}, {Manera}, {Monaco}, {Murray}, {Knebe},
  {Scoccola}, {Yepes}, {Garcia-Bellido}, {Marin}, {Muller}, {Skibba}, {Crocce},
  {Fosalba}, {Gottlober}, {Klypin}, {Power}, {Tao}, \&
  {Turchaninov}}]{paper:chuang2014}
{Chuang} C.-H. {et~al.}, 2014, ArXiv e-prints, arXiv:1412.7729

\bibitem[{{Coles} \& {Jones}(1991)}]{paper:coles1991}
{Coles} P., {Jones} B., 1991, \mnras, 248, 1

\bibitem[{{Covone} {et~al}\mbox{.}(2014){Covone}, {Sereno}, {Kilbinger}, \&
  {Cardone}}]{paper:covone2014}
{Covone} G., {Sereno} M., {Kilbinger} M., {Cardone} V.~F., 2014, \apjl, 784,
  L25

\bibitem[{{Cuesta} {et~al}\mbox{.}(2015){Cuesta}, {Vargas-Maga{\~n}a},
  {Beutler}, {Bolton}, {Brownstein}, {Eisenstein}, {Gil-Mar{\'{\i}}n}, {Ho},
  {McBride}, {Maraston}, {Padmanabhan}, {Percival}, {Reid}, {Ross}, {Ross},
  {S{\'a}nchez}, {Schlegel}, {Schneider}, {Thomas}, {Tinker}, {Tojeiro},
  {Verde}, \& {White}}]{paper:cuesta2015}
{Cuesta} A.~J. {et~al.}, 2015, ArXiv e-prints, arXiv:1509.06371

\bibitem[{{Dawson} {et~al}\mbox{.}(2015){Dawson}, {Kneib}, {Percival}, {Alam},
  {Albareti}, {Anderson}, {Armengaud}, {Aubourg}, {Bailey}, {Bautista},
  {Berlind}, {Bershady}, {Beutler}, {Bizyaev}, {Blanton}, {Blomqvist},
  {Bolton}, {Bovy}, {Brandt}, {Brinkmann}, {Brownstein}, {Burtin}, {Busca},
  {Cai}, {Chuang}, {Clerc}, {Comparat}, {Cope}, {Croft}, {Cruz-Gonzalez}, {da
  Costa}, {Cousinou}, {Darling}, {de la Torre}, {Delubac}, {du Mas des
  Bourboux}, {Dwelly}, {Ealet}, {Eisenstein}, {Eracleous}, {Escoffier}, {Fan},
  {Finoguenov}, {Font-Ribera}, {Frinchaboy}, {Gaulme}, {Georgakakis}, {Green},
  {Guo}, {Guy}, {Ho}, {Holder}, {Huehnerhoff}, {Hutchinson}, {Jing}, {Jullo},
  {Kamble}, {Kinemuchi}, {Kirkby}, {Kitaura}, {Klaene}, {Laher}, {Lang},
  {Laurent}, {Le Goff}, {Li}, {Liang}, {Lima}, {Lin}, {Lin}, {Lin}, {Long},
  {Lundgren}, {MacDonald}, {Geimba Maia}, {Malanushenko}, {Malanushenko},
  {Mariappan}, {McBride}, {McGreer}, {Menard}, {Merloni}, {Meza},
  {Montero-Dorta}, {Muna}, {Myers}, {Nandra}, {Naugle}, {Newman}, {Noterdaeme},
  {Nugent}, {Ogando}, {Olmstead}, {Oravetz}, {Oravetz}, {Padmanabhan},
  {Palanque-Delabrouille}, {Pan}, {Parejko}, {Paris}, {Peacock}, {Petitjean},
  {Pieri}, {Pisani}, {Prada}, {Prakash}, {Raichoor}, {Reid}, {Rich}, {Ridl},
  {Rodriguez-Torres}, {Carnero Rosell}, {Ross}, {Rossi}, {Ruan}, {Salvato},
  {Sayres}, {Schneider}, {Schlegel}, {Seljak}, {Seo}, {Sesar}, {Shandera},
  {Shu}, {Slosar}, {Sobreira}, {Strauss}, {Streblyanska}, {Suzuki}, {Tao},
  {Tinker}, {Tojeiro}, {Vargas-Magana}, {Wang}, {Weaver}, {Weinberg}, {White},
  {Wood-Vasey}, {Yeche}, {Zhai}, {Zhao}, {Zhao}, {Zheng}, {Ben Zhu}, \&
  {Zou}}]{paper:dawson2015}
{Dawson} K.~S. {et~al.}, 2015, ArXiv e-prints, arXiv:1508.04473

\bibitem[{{Dawson} {et~al}\mbox{.}(2013){Dawson}, {Schlegel}, {Ahn},
  {Anderson}, {Aubourg}, {Bailey}, {Barkhouser}, {Bautista}, {Beifiori},
  {Berlind}, {Bhardwaj}, {Bizyaev}, {Blake}, {Blanton}, {Blomqvist}, {Bolton},
  {Borde}, {Bovy}, {Brandt}, {Brewington}, {Brinkmann}, {Brown}, {Brownstein},
  {Bundy}, {Busca}, {Carithers}, {Carnero}, {Carr}, {Chen}, {Comparat},
  {Connolly}, {Cope}, {Croft}, {Cuesta}, {da Costa}, {Davenport}, {Delubac},
  {de Putter}, {Dhital}, {Ealet}, {Ebelke}, {Eisenstein}, {Escoffier}, {Fan},
  {Filiz Ak}, {Finley}, {Font-Ribera}, {G{\'e}nova-Santos}, {Gunn}, {Guo},
  {Haggard}, {Hall}, {Hamilton}, {Harris}, {Harris}, {Ho}, {Hogg}, {Holder},
  {Honscheid}, {Huehnerhoff}, {Jordan}, {Jordan}, {Kauffmann}, {Kazin},
  {Kirkby}, {Klaene}, {Kneib}, {Le Goff}, {Lee}, {Long}, {Loomis}, {Lundgren},
  {Lupton}, {Maia}, {Makler}, {Malanushenko}, {Malanushenko}, {Mandelbaum},
  {Manera}, {Maraston}, {Margala}, {Masters}, {McBride}, {McDonald}, {McGreer},
  {McMahon}, {Mena}, {Miralda-Escud{\'e}}, {Montero-Dorta}, {Montesano},
  {Muna}, {Myers}, {Naugle}, {Nichol}, {Noterdaeme}, {Nuza}, {Olmstead},
  {Oravetz}, {Oravetz}, {Owen}, {Padmanabhan}, {Palanque-Delabrouille}, {Pan},
  {Parejko}, {P{\^a}ris}, {Percival}, {P{\'e}rez-Fournon},
  {P{\'e}rez-R{\`a}fols}, {Petitjean}, {Pfaffenberger}, {Pforr}, {Pieri},
  {Prada}, {Price-Whelan}, {Raddick}, {Rebolo}, {Rich}, {Richards}, {Rockosi},
  {Roe}, {Ross}, {Ross}, {Rossi}, {Rubi{\~n}o-Martin}, {Samushia},
  {S{\'a}nchez}, {Sayres}, {Schmidt}, {Schneider}, {Sc{\'o}ccola}, {Seo},
  {Shelden}, {Sheldon}, {Shen}, {Shu}, {Slosar}, {Smee}, {Snedden}, {Stauffer},
  {Steele}, {Strauss}, {Streblyanska}, {Suzuki}, {Swanson}, {Tal}, {Tanaka},
  {Thomas}, {Tinker}, {Tojeiro}, {Tremonti}, {Vargas Maga{\~n}a}, {Verde},
  {Viel}, {Wake}, {Watson}, {Weaver}, {Weinberg}, {Weiner}, {West}, {White},
  {Wood-Vasey}, {Yeche}, {Zehavi}, {Zhao}, \& {Zheng}}]{paper:boss}
{Dawson} K.~S. {et~al.}, 2013, \aj, 145, 10

\bibitem[{{Delubac} {et~al}\mbox{.}(2015){Delubac}, {Bautista}, {Busca},
  {Rich}, {Kirkby}, {Bailey}, {Font-Ribera}, {Slosar}, {Lee}, {Pieri},
  {Hamilton}, {Aubourg}, {Blomqvist}, {Bovy}, {Brinkmann}, {Carithers},
  {Dawson}, {Eisenstein}, {Gontcho}, {Kneib}, {Le Goff}, {Margala},
  {Miralda-Escud{\'e}}, {Myers}, {Nichol}, {Noterdaeme}, {O'Connell},
  {Olmstead}, {Palanque-Delabrouille}, {P{\^a}ris}, {Petitjean}, {Ross},
  {Rossi}, {Schlegel}, {Schneider}, {Weinberg}, {Y{\`e}che}, \&
  {York}}]{paper:delubac2015}
{Delubac} T. {et~al.}, 2015, \aap, 574, A59

\bibitem[{{Eisenstein} \& {Hu}(1998)}]{paper:eisenstein1998}
{Eisenstein} D.~J., {Hu} W., 1998, \apj, 496, 605

\bibitem[{{Eisenstein}, {Seo} \& {White}(2007){Eisenstein}, {Seo}, \&
  {White}}]{paper:eisenstein2007}
{Eisenstein} D.~J., {Seo} H.-J., {White} M., 2007, \apj, 664, 660

\bibitem[{{Estrada}, {Sefusatti} \& {Frieman}(2009){Estrada}, {Sefusatti}, \&
  {Frieman}}]{paper:estrada2009}
{Estrada} J., {Sefusatti} E., {Frieman} J.~A., 2009, \apj, 692, 265

\bibitem[{{Hong} {et~al}\mbox{.}(2012){Hong}, {Han}, {Wen}, {Sun}, \&
  {Zhan}}]{paper:hong2012}
{Hong} T., {Han} J.~L., {Wen} Z.~L., {Sun} L., {Zhan} H., 2012, \apj, 749, 81

\bibitem[{{H{\"u}tsi}(2010)}]{paper:hutsi2010}
{H{\"u}tsi} G., 2010, \mnras, 401, 2477

\bibitem[{{Kazin} {et~al}\mbox{.}(2014){Kazin}, {Koda}, {Blake}, {Padmanabhan},
  {Brough}, {Colless}, {Contreras}, {Couch}, {Croom}, {Croton}, {Davis},
  {Drinkwater}, {Forster}, {Gilbank}, {Gladders}, {Glazebrook}, {Jelliffe},
  {Jurek}, {Li}, {Madore}, {Martin}, {Pimbblet}, {Poole}, {Pracy}, {Sharp},
  {Wisnioski}, {Woods}, {Wyder}, \& {Yee}}]{paper:kazin2014}
{Kazin} E.~A. {et~al.}, 2014, \mnras, 441, 3524

\bibitem[{{Landy} \& {Szalay}(1993)}]{paper:landy1993}
{Landy} S.~D., {Szalay} A.~S., 1993, \apj, 412, 64

\bibitem[{{Laureijs} {et~al}\mbox{.}(2011){Laureijs}, {Amiaux}, {Arduini},
  {Augu{\`e}res}, {Brinchmann}, {Cole}, {Cropper}, {Dabin}, {Duvet}, {Ealet},
  \& et~al.}]{paper:laureijs2011}
{Laureijs} R. {et~al.}, 2011, ArXiv e-prints, arXiv:1110.3193

\bibitem[{{Lewis} \& {Bridle}(2002)}]{paper:lewis2002}
{Lewis} A., {Bridle} S., 2002, \prd, 66, 103511

\bibitem[{{Mana} {et~al}\mbox{.}(2013){Mana}, {Giannantonio}, {Weller},
  {Hoyle}, {H{\"u}tsi}, \& {Sartoris}}]{paper:mana2013}
{Mana} A., {Giannantonio} T., {Weller} J., {Hoyle} B., {H{\"u}tsi} G.,
  {Sartoris} B., 2013, \mnras, 434, 684

\bibitem[{{Marulli} {et~al}\mbox{.}(2012){Marulli}, {Bianchi}, {Branchini},
  {Guzzo}, {Moscardini}, \& {Angulo}}]{paper:marulli2012}
{Marulli} F., {Bianchi} D., {Branchini} E., {Guzzo} L., {Moscardini} L.,
  {Angulo} R.~E., 2012, \mnras, 426, 2566

\bibitem[{{Marulli}, {Veropalumbo} \& {Moresco}(in preparation){Marulli},
  {Veropalumbo}, \& {Moresco}}]{paper:marulli2015}
{Marulli} F., {Veropalumbo} A., {Moresco} M., in preparation

\bibitem[{{Marulli} {et~al}\mbox{.}(2015){Marulli}, {Veropalumbo},
  {Moscardini}, {Cimatti}, \& {Dolag}}]{paper:marulli2015b}
{Marulli} F., {Veropalumbo} A., {Moscardini} L., {Cimatti} A., {Dolag} K.,
  2015, ArXiv e-prints, arXiv:1505.01170

\bibitem[{{Merloni} {et~al}\mbox{.}(2012){Merloni}
  {et~al.}}]{paper:merloni2012}
{Merloni} A., {et~al.}, 2012, ArXiv e-prints, arXiv:1209.3114

\bibitem[{{Norberg} {et~al}\mbox{.}(2009){Norberg}, {Baugh}, {Gazta{\~n}aga},
  \& {Croton}}]{paper:norberg2009}
{Norberg} P., {Baugh} C.~M., {Gazta{\~n}aga} E., {Croton} D.~J., 2009, \mnras,
  396, 19

\bibitem[{{Planck Collaboration} {et~al}\mbox{.}(2014){Planck Collaboration},
  {Ade}, {Aghanim}, {Armitage-Caplan}, {Arnaud}, {Ashdown}, {Atrio-Barandela},
  {Aumont}, {Baccigalupi}, {Banday}, \& et~al.}]{paper:planck2013}
{Planck Collaboration} {et~al.}, 2014, \aap, 571, A16

\bibitem[{{Ross} {et~al}\mbox{.}(2012){Ross}, {Percival}, {S{\'a}nchez},
  {Samushia}, {Ho}, {Kazin}, {Manera}, {Reid}, {White}, {Tojeiro}, {McBride},
  {Xu}, {Wake}, {Strauss}, {Montesano}, {Swanson}, {Bailey}, {Bolton}, {Dorta},
  {Eisenstein}, {Guo}, {Hamilton}, {Nichol}, {Padmanabhan}, {Prada},
  {Schlegel}, {Maga{\~n}a}, {Zehavi}, {Blanton}, {Bizyaev}, {Brewington},
  {Cuesta}, {Malanushenko}, {Malanushenko}, {Oravetz}, {Parejko}, {Pan},
  {Schneider}, {Shelden}, {Simmons}, {Snedden}, \& {Zhao}}]{paper:ross2012}
{Ross} A.~J. {et~al.}, 2012, \mnras, 424, 564

\bibitem[{{Ross} {et~al}\mbox{.}(2015){Ross}, {Samushia}, {Howlett},
  {Percival}, {Burden}, \& {Manera}}]{paper:ross2015}
{Ross} A.~J., {Samushia} L., {Howlett} C., {Percival} W.~J., {Burden} A.,
  {Manera} M., 2015, \mnras, 449, 835

\bibitem[{{S{\'a}nchez} {et~al}\mbox{.}(2013){S{\'a}nchez}, {Kazin}, {Beutler},
  {Chuang}, {Cuesta}, {Eisenstein}, {Manera}, {Montesano}, {Nichol},
  {Padmanabhan}, {Percival}, {Prada}, {Ross}, {Schlegel}, {Tinker}, {Tojeiro},
  {Weinberg}, {Xu}, {Brinkmann}, {Brownstein}, {Schneider}, \&
  {Thomas}}]{paper:sanchez2013}
{S{\'a}nchez} A.~G. {et~al.}, 2013, \mnras, 433, 1202

\bibitem[{{Sartoris} {et~al}\mbox{.}(2015){Sartoris}, {Biviano}, {Fedeli},
  {Bartlett}, {Borgani}, {Costanzi}, {Giocoli}, {Moscardini}, {Weller},
  {Ascaso}, {Bardelli}, {Maurogordato}, \& {Viana}}]{paper:sartoris2015}
{Sartoris} B. {et~al.}, 2015, ArXiv e-prints, arXiv:1505.02165

\bibitem[{{Sereno} {et~al}\mbox{.}(2015){Sereno}, {Veropalumbo}, {Marulli},
  {Covone}, {Moscardini}, \& {Cimatti}}]{paper:sereno2014}
{Sereno} M., {Veropalumbo} A., {Marulli} F., {Covone} G., {Moscardini} L.,
  {Cimatti} A., 2015, \mnras, 449, 4147

\bibitem[{{Swanson} {et~al}\mbox{.}(2008){Swanson}, {Tegmark}, {Hamilton}, \&
  {Hill}}]{paper:swanson2008}
{Swanson} M.~E.~C., {Tegmark} M., {Hamilton} A.~J.~S., {Hill} J.~C., 2008,
  \mnras, 387, 1391

\bibitem[{{Tempel} {et~al}\mbox{.}(2014){Tempel}, {Tamm}, {Gramann},
  {Tuvikene}, {Liivam{\"a}gi}, {Suhhonenko}, {Kipper}, {Einasto}, \&
  {Saar}}]{paper:tempel2014}
{Tempel} E. {et~al.}, 2014, \aap, 566, A1

\bibitem[{{Veropalumbo} {et~al}\mbox{.}(2014){Veropalumbo}, {Marulli},
  {Moscardini}, {Moresco}, \& {Cimatti}}]{paper:veropalumbo2014}
{Veropalumbo} A., {Marulli} F., {Moscardini} L., {Moresco} M., {Cimatti} A.,
  2014, \mnras, 442, 3275

\bibitem[{{Wen}, {Han} \& {Liu}(2009){Wen}, {Han}, \& {Liu}}]{paper:wen2009}
{Wen} Z.~L., {Han} J.~L., {Liu} F.~S., 2009, \apjs, 183, 197

\bibitem[{{Wen}, {Han} \& {Liu}(2012){Wen}, {Han}, \& {Liu}}]{paper:whl2012}
{Wen} Z.~L., {Han} J.~L., {Liu} F.~S., 2012, \apjs, 199, 34

\end{thebibliography}

\label{lastpage}
\end{document}